\documentclass[english,aip,reprint]{revtex4-1}
\usepackage{mathptmx}
\usepackage{helvet}
\usepackage[LGR,T1]{fontenc}
\usepackage[latin9]{inputenc}
\usepackage{color}
\usepackage{babel}
\usepackage{array}
\usepackage{booktabs}
\usepackage{textcomp}
\usepackage{multirow}
\usepackage{amsmath}
\usepackage{graphicx}
\usepackage{subscript}
\usepackage[unicode=true,pdfusetitle,
 bookmarks=true,bookmarksnumbered=false,bookmarksopen=false,
 breaklinks=false,pdfborder={0 0 1},backref=false,colorlinks=true]
 {hyperref}
\hypersetup{
 urlcolor=blue,linkcolor=red,citecolor=blue}

\makeatletter

\DeclareRobustCommand{\greektext}{%
  \fontencoding{LGR}\selectfont\def\encodingdefault{LGR}}
\DeclareRobustCommand{\textgreek}[1]{\leavevmode{\greektext #1}}
\ProvideTextCommand{\~}{LGR}[1]{\char126#1}

\newcommand{\lyxmathsym}[1]{\ifmmode\begingroup\def\b@ld{bold}
  \text{\ifx\math@version\b@ld\bfseries\fi#1}\endgroup\else#1\fi}

\providecommand{\tabularnewline}{\\}

\makeatother

\begin{document}
\title{Molecular beam epitaxy of single-crystalline bixbyite (In\textsubscript{1-\textit{x}}Ga\textsubscript{\textit{x}})\textsubscript{2}O\textsubscript{3}
films (\textit{x}\ensuremath{\le}0.18): Structural properties and
consequences of compositional inhomogeneity }
\author{Alexandra Papadogianni}
\affiliation{Paul-Drude-Institut für Festkörperelektronik, Leibniz-Institut im
Forschungsverbund Berlin e.V., Hausvogteiplatz 5\textendash 7, 10117
Berlin, Germany}
\author{Charlotte Wouters}
\affiliation{Leibniz-Institut für Kristallzüchtung, Max-Born-Str. 2, 12489 Berlin,
Germany}
\author{Robert Schewski}
\affiliation{Leibniz-Institut für Kristallzüchtung, Max-Born-Str. 2, 12489 Berlin,
Germany}
\author{Johannes Feldl}
\affiliation{Paul-Drude-Institut für Festkörperelektronik, Leibniz-Institut im
Forschungsverbund Berlin e.V., Hausvogteiplatz 5\textendash 7, 10117
Berlin, Germany}
\author{Jonas Lähnemann}
\affiliation{Paul-Drude-Institut für Festkörperelektronik, Leibniz-Institut im
Forschungsverbund Berlin e.V., Hausvogteiplatz 5\textendash 7, 10117
Berlin, Germany}
\author{\textcolor{black}{Takahiro Nagata}}
\affiliation{\textcolor{black}{National Institute for Materials Science, 1-1 Namiki
Tsukuba, 305-0044 Ibaraki, Japan}}
\author{Elias Kluth}
\affiliation{Institut für Experimentelle Physik, Otto-von-Guericke-Universität
Magdeburg, Universitätsplatz 2, 39106 Magdeburg, Germany}
\author{Martin Feneberg}
\affiliation{Institut für Experimentelle Physik, Otto-von-Guericke-Universität
Magdeburg, Universitätsplatz 2, 39106 Magdeburg, Germany}
\author{Rüdiger Goldhahn}
\affiliation{Institut für Experimentelle Physik, Otto-von-Guericke-Universität
Magdeburg, Universitätsplatz 2, 39106 Magdeburg, Germany}
\author{Manfred Ramsteiner}
\affiliation{Paul-Drude-Institut für Festkörperelektronik, Leibniz-Institut im
Forschungsverbund Berlin e.V., Hausvogteiplatz 5\textendash 7, 10117
Berlin, Germany}
\author{Martin Albrecht}
\affiliation{Leibniz-Institut für Kristallzüchtung, Max-Born-Str. 2, 12489 Berlin,
Germany}
\author{Oliver Bierwagen}
\affiliation{Paul-Drude-Institut für Festkörperelektronik, Leibniz-Institut im
Forschungsverbund Berlin e.V., Hausvogteiplatz 5\textendash 7, 10117
Berlin, Germany}
\date{\today}
\begin{abstract}
In this work, we show the heteroepitaxial growth of single-crystalline
bixbyite (In\textsubscript{1-\textit{x}}Ga\textsubscript{\textit{x}})\textsubscript{2}O\textsubscript{3}
films on (111)-oriented yttria-stabilized zirconia substrates using
plasma-assisted molecular beam epitaxy under various growth conditions.
A pure In\textsubscript{2}O\textsubscript{3} buffer layer between
the substrate and (In\textsubscript{1-\textit{x}}Ga\textsubscript{\textit{x}})\textsubscript{2}O\textsubscript{3}
alloy is shown to result in smoother film surfaces and significantly
improved crystallinity. Symmetric out-of-plane $2\theta\lyxmathsym{\textendash}\omega$
x-ray diffraction scans show a single (111) crystal orientation and
transmission electron microscopy confirms the single-crystallinity
up to $x=0.18$ and only slight film quality deterioration with increasing
Ga content. Partially relaxed layers are demonstrated via reciprocal
space mapping with lattice parameters fitting well to Vegard's law.
However, the Ga cations are not evenly distributed within the films
containing nominally $x>0.11$: inclusions with high Ga density up
to $x=0.50$ are observed within a \textquotedbl matrix\textquotedbl{}
with $x\approx0.08$. The cubic bixbyite phase is preserved, in both
the \textquotedbl matrix\textquotedbl{} and the inclusions. \textcolor{black}{Moreover,
for }$x\geq0.11$\textcolor{black}{, both the Raman phonon lines as
well as the optical absorption onset remain nearly constant.} \textcolor{black}{Hard
x-ray photoelectron spectroscopy measurements also indicate a widening
of the band gap and exhibit similar saturation of the Ga 2}\textit{\textcolor{black}{p}}\textcolor{black}{{}
core level position for high Ga contents. This saturation behavior
of the spectroscopic properties further supports the limited Ga incorporation
into the \textquotedbl matrix\textquotedbl{} of the film.}
\end{abstract}
\maketitle

\section{Introduction}

The group-III sesquioxides In\textsubscript{2}O\textsubscript{3}
and Ga\textsubscript{2}O\textsubscript{3} are transparent semiconducting
materials widely studied both for their implementation in electronic
devices and\textemdash from a fundamental semiconductor physics perspective\textemdash for
their unique properties. Although these two oxide compounds are isovalent
and have the same chemical configuration, the equilibrium structures
of In\textsubscript{2}O\textsubscript{3} and Ga\textsubscript{2}O\textsubscript{3 }are
very different, which can be explained in terms of Coulomb and orbital
interactions, influenced by cation size and valence electron energies
respectively.\citep{Ma_Wei_2015}

On the one hand, In\textsubscript{2}O\textsubscript{3} has a stable
cubic bixbyite crystal structure and belongs to the space group 206
($Ia\bar{3}$) with a lattice parameter of $a_{\mathrm{In_{2}O_{3}}}=10.117\,\mathrm{\mathring{A}}$,\citep{Marezio1966}
as well as a less-studied metastable rhombohedral structure of the
space group 167 ($R\bar{3}c$). Bixbyite In\textsubscript{2}O\textsubscript{3}
possesses an optically forbidden \textcolor{red}{}direct band gap
of approximately 2.7\textendash 2.9~eV, with strong optical absorption
occurring from valence bands nearly 1~eV below the valence band maximum
(VBM),\citep{Walsh_PRL.100.167402,King_PRB.79.205211,Irmscher_10.1002/pssa.201330184}\textcolor{black}{{}
resulting in an onset of strong optical absorption at around 3.7~eV.}
This property of In\textsubscript{2}O\textsubscript{3} renders it
transparent in the visible regime and is remarkably combined with
high electrical conductivity; as In\textsubscript{2}O\textsubscript{3}
exhibits inherent \textit{n}-type conductivity, due to what is commonly
referred to as unintentional doping. This unintentional conductivity
can be significantly enhanced by intentional donor doping. Sn-doped
In\textsubscript{2}O\textsubscript{3}\textemdash commonly known
as indium-tin oxide (ITO)\citep{chae2001,tiwari2004,tsai_2016}\textemdash is
the most widely commercially used transparent conducting oxide (TCO),
as a transparent contact in optoelectronics, such as displays, light-emitting
diodes, and solar cells.

On the other hand, Ga\textsubscript{2}O\textsubscript{3}, which
has been under the spotlight for the past few years, attracting interest
for applications in ultraviolet (UV) photodetectors and high-power
devices. It has several polymorphs,\citep{Roy_JACS.10.1021/ja01123a039,Galazka_2018}
the most stable out of which is its $\beta$-phase. \textit{$\beta$}-Ga\textsubscript{2}O\textsubscript{3}
has a monoclinic crystal structure, which belongs to the space group
12 (\textit{C}2/\textit{m}), and possesses a band gap of approximately
4.8~eV.\citep{Matsumoto_1974} It is, thus, transparent within both
the visible and well into the UV range. \textcolor{black}{A bixbyite
``$\delta$''-}Ga\textsubscript{2}O\textsubscript{3}\textcolor{black}{{}
phase has been theoretically predicted to have a lattice parameter
of} approximately $a_{\mathrm{Ga_{2}O_{3}}}=9.190-9.401\,\mathrm{\mathring{A}}$\citep{peelaers2015,Yoshioka_2007}
and a band gap of 5.0~eV, according to the latest report.\citep{peelaers2015}
To our knowledge, this phase has not been experimentally demonstrated
yet,\citep{Playford_2013} despite the early reports of \citet[.][]{Roy_JACS.10.1021/ja01123a039}
Finally, contrary to In\textsubscript{2}O\textsubscript{3}, high-quality,
pure Ga\textsubscript{2}O\textsubscript{3} films are insulating
at room temperature.\citep{Wong_JpnJApplPhys55-2016}

A combination of these two oxides allows for adjustments of the properties
of the two original compounds and can, for instance, facilitate band
gap engineering,\citep{Hill_1974} as observed in other semiconductor
systems. Moreover, concerning potential applications, the incorporation
of Ga in the In\textsubscript{2}O\textsubscript{3} crystal lattice
would potentially provide us with a wide-band-gap TCO, assuming that
additional doping with Sn can be achieved. Several studies have been
done towards this direction, with films prepared by various techniques
covering the entire composition range of (In\textsubscript{1-\textit{x}}Ga\textsubscript{\textit{x}})\textsubscript{2}O\textsubscript{3},\citep{kokubun_2010}
though most of the works found in the literature focus on the high-\textit{x},
Ga-rich end.\citep{Vasyltsiv_PSSB(1996),Oshima_PSSC_2008,kranert_2014,baldini_2015}
Due to the different crystal structures of the parent materials, structural
changes over the composition range are expected, which should limit
their miscibility and affect the electronic properties of the resulting
alloy.\textcolor{blue}{{} }\textcolor{black}{Density-functional theory
calculations estimate that the miscibility gap opens at $x=0.15$
for free-standing (In}\textsubscript{\textcolor{black}{1-}\textit{\textcolor{black}{x}}}\textcolor{black}{Ga}\textsubscript{\textit{\textcolor{black}{x}}}\textcolor{black}{)}\textsubscript{\textcolor{black}{2}}\textcolor{black}{O}\textsubscript{\textcolor{black}{3}}\textcolor{black}{{}
and at $x=0.45$ for the bixbyite alloy grown epitaxially on In}\textsubscript{\textcolor{black}{2}}\textcolor{black}{O}\textsubscript{\textcolor{black}{3}}\textcolor{black}{,
independent of the growth temperature.\citep{Maccioni_2016}}\textcolor{blue}{{}
}\textcolor{black}{The recent work of \citet{Wouters_IGO_2020} predicts
the cubic phase to be stable up to approximately $x=0.10$. Previous
experimental investigations on (In}\textsubscript{\textcolor{black}{1-}\textit{\textcolor{black}{x}}}\textcolor{black}{Ga}\textsubscript{\textit{\textcolor{black}{x}}}\textcolor{black}{)}\textsubscript{\textcolor{black}{2}}\textcolor{black}{O}\textsubscript{\textcolor{black}{3}}\textcolor{black}{{}
polycrystalline ceramic alloys\citep{regoutz_2015} identify the solubility
limit of Ga at approximately $x=0.10$, but, surprisingly, report
a decrease of the In}\textsubscript{\textcolor{black}{2}}\textcolor{black}{O}\textsubscript{\textcolor{black}{3}}\textcolor{black}{{}
band gap with Ga content.}\textcolor{blue}{{} }\textcolor{black}{Another
preceding experimental study on films grown by metalorganic chemical
vapor deposition\citep{kong_2010} demonstrates a bixbyite phase for
Ga contents up to $x=0.50$ and a shift of the optical absorption
edge towards shorter wavelengths, which indicates a widening of the
band gap. However, it is unclear whether the films are homogeneous
or even contain amorphous regions. Recently \citet{Swallow_2021}
have clearly demonstrated valence band edge shift away from the Fermi
level, increase in optical gap, and depletion of surface space charge
with increasing Ga content in textured (In}\textsubscript{\textcolor{black}{1-}\textit{\textcolor{black}{x}}}\textcolor{black}{Ga}\textsubscript{\textit{\textcolor{black}{x}}}\textcolor{black}{)}\textsubscript{\textcolor{black}{2}}\textcolor{black}{O}\textsubscript{\textcolor{black}{3}}\textcolor{black}{{}
films grown by pulsed layer deposition. \citet{Nagata_2020} also
recently published a combinatorial thin film synthesis study showing
single-crystalline bixbyite (In}\textsubscript{\textcolor{black}{1-}\textit{\textcolor{black}{x}}}\textcolor{black}{Ga}\textsubscript{\textit{\textcolor{black}{x}}}\textcolor{black}{)}\textsubscript{\textcolor{black}{2}}\textcolor{black}{O}\textsubscript{\textcolor{black}{3}}\textcolor{black}{{}
layers with a low Ga composition. Nonetheless, there is still }a lack
of literature with detailed structural information on well-defined
high-quality \textcolor{black}{(In}\textsubscript{\textcolor{black}{1-}\textit{\textcolor{black}{x}}}\textcolor{black}{Ga}\textsubscript{\textit{\textcolor{black}{x}}}\textcolor{black}{)}\textsubscript{\textcolor{black}{2}}\textcolor{black}{O}\textsubscript{\textcolor{black}{3}}\textcolor{black}{{}
films} with low $x$ allowing for deep exploration of the fundamental
physical properties of the bixbyite phase of this alloy.

In this work, we synthesize single-crystalline phase-pure thin (In\textsubscript{1-\textit{x}}Ga\textsubscript{\textit{x}})\textsubscript{2}O\textsubscript{3}
films up to $x=0.18$ using plasma-assisted molecular beam epitaxy
(PA-MBE) and investigate the effect of a pure In\textsubscript{2}O\textsubscript{3}
buffer layer and substrate temperature\textemdash especially during
nucleation\textemdash on the film quality. The resulting films are
evaluated in terms of crystallinity, lattice parameter, compositional
homogeneity, optical band gap, and Raman phonon modes shift with Ga
content.

\section{Experimental details}

\paragraph*{Substrate choice and preparation}

For the purposes of this study, high quality (111)-oriented single-crystalline
(In\textsubscript{1-\textit{x}}Ga\textsubscript{\textit{x}})\textsubscript{2}O\textsubscript{3}
films have been synthesized by PA-MBE on quarters of 2-in. insulating
ZrO\textsubscript{2}:Y (YSZ) (111) substrates. The study focuses
on the low-\textit{x} bixbyite phase end of (In\textsubscript{1-\textit{x}}Ga\textsubscript{\textit{x}})\textsubscript{2}O\textsubscript{3},
hence the substrate choice is based on its suitability for heteroepitaxy
of pure In\textsubscript{2}O\textsubscript{3}. Along the {[}100{]}
direction In\textsubscript{2}O\textsubscript{3} crystallizes on
a $2\times2$ YSZ unit cell with a cube-on-cube epitaxial relation\citep{tarsa_1993}
with a low tensile mismatch of 1.6-1.7\,\%,\citep{Cowley_PhysRevB.82.165312,Bierwagen_doi:10.1063/1.3276910}
as In\textsubscript{2}O\textsubscript{3} has a lattice constant
of 10.117~Å\citep{Marezio1966}\textemdash slightly smaller than
twice that of YSZ, for a Y concentration of 10\,mol\%\citep{HAYASHI2005613}
(as the ones used in this study), which amounts to approximately 10.28~Å.
The choice to grow the (111) surface of In\textsubscript{2}O\textsubscript{3}
was based on the fact that the (111) plane has the lowest surface
tension, i.e., surface free energy per unit area, compared to the
other low-index surfaces of In\textsubscript{2}O\textsubscript{3},\citep{Agoston_PhysRevB.84.045311}
which facilitates the growth of smooth, unfaceted films.\citep{Bierwagen_faceting_2016}
\textcolor{black}{Prior to loading into the PA-MBE system, the substrates
were cleaned in an ultrasonic bath with organic solvents (butyl acetate,
acetone, and isopropanol for 5 minutes in each) and, finally, rinsed
off with deionized water.} In addition, the substrates were thermally
cleaned in the MBE growth chamber at $800\thinspace\mathrm{{^\circ}C}$
for 10~min under a high oxygen plasma flux of 1~standard cubic centimeter
per minute (SCCM). The power of the oxygen plasma source (6N purity)
was maintained at 300~W throughout both the annealing and growth
procedures.

\paragraph*{Film growth and in-situ film characterization}

Table~\ref{growth_details} summarizes the growth parameters used
for the growth of the samples investigated in this study.
\begin{table*}
\caption{Growth parameters for (In\protect\textsubscript{1-\textit{x}}Ga\protect\textsubscript{\textit{x}})\protect\textsubscript{2}O\protect\textsubscript{3}
samples on YSZ (111): $T_{\mathrm{nucl}}$ denotes the substrate temperature
during the nucleation process of the pure In\protect\textsubscript{2}O\protect\textsubscript{3}
buffer layer (with the exception of I.b, where Ga was also provided),
$T_{\mathrm{max}}$ is the maximum substrate temperature reached during
the smoothing process, $T_{\mathrm{growth}}$ is the final growth
temperature, $t_{\mathrm{nucl}}$ and $t_{\mathrm{growth}}$ are the
growth times for the nucleation layer and the bulk of the film respectively,
O-flux corresponds to the oxygen flux used for both the nucleation
and growth and the value in brackets for series III to the increased
value during the growth interruption and surface smoothing, BEP\protect\textsubscript{In/Ga}
is the beam equivalent pressure of the respective metal cell, $d_{\mathrm{film}}$
the total thickness of the resulting film (including the buffer layer),
$x_{\mathrm{EDX}}$ is the corresponding Ga cation content\textemdash given
in cation percentage, as measured on the as-grown samples by energy-dispersive
X-ray spectroscopy (EDX). $R_{\mathrm{q}}^{\mathrm{10\,\mu m}}$ is
the root mean square (RMS) roughness of a $10\times10\,\mathrm{\mu m^{2}}$
AFM image and $\omega_{\mathrm{FWHM}}^{222}$ is the full width at
half maximum (FWHM) of the rocking curve of the (In,Ga)\protect\textsubscript{2}O\protect\textsubscript{3}
222 peak.}

\begin{tabular}{cccccccccccccccc}
\toprule 
\textcolor{black}{Series} & \textcolor{black}{Sample} & \textcolor{black}{$T_{\mathrm{nucl}}$} & \textcolor{black}{$T_{\mathrm{max}}$} & \textcolor{black}{$T_{\mathrm{growth}}$} & \textcolor{black}{$t_{\mathrm{nucl}}$} & \textcolor{black}{$t_{\mathrm{growth}}$} & \textcolor{black}{O-flux} & \textcolor{black}{BEP}\textsubscript{\textcolor{black}{In}} & \textcolor{black}{BEP}\textsubscript{\textcolor{black}{Ga}} & \textcolor{black}{BEP}\textsubscript{\textcolor{black}{Sn}} & \textcolor{black}{$d_{\mathrm{film}}$} & \textcolor{black}{$x_{\mathrm{EDX}}$} & In\textsubscript{2}O\textsubscript{3} & \textcolor{black}{$R_{\mathrm{q}}^{\mathrm{10\,\mu m}}$} & \textcolor{black}{$\omega_{\mathrm{FWHM}}^{222}$}\tabularnewline
 &  & \textcolor{black}{($\mathrm{{^\circ}C}$ )} & \textcolor{black}{($\mathrm{{^\circ}C}$ )} & \textcolor{black}{($\mathrm{{^\circ}C}$ )} & \textcolor{black}{(s)} & \textcolor{black}{(s)} & \textcolor{black}{(SCCM)} & \textcolor{black}{($10^{-7}$ mbar)} & \textcolor{black}{($10^{-8}$ mbar)} & \textcolor{black}{($10^{-8}$ mbar)} & \textcolor{black}{(nm)} & \textcolor{black}{(cat.)} & buffer & \textcolor{black}{(nm)} & \textcolor{black}{(°)}\tabularnewline
\midrule
\midrule 
\multirow{3}{*}{\textcolor{black}{I}} & \textcolor{black}{a} & \multirow{3}{*}{\textcolor{black}{600}} & \multirow{3}{*}{\textcolor{black}{750}} & \multirow{3}{*}{\textcolor{black}{750}} & \multirow{3}{*}{\textcolor{black}{270}} & \multirow{3}{*}{\textcolor{black}{2400}} & \multirow{3}{*}{\textcolor{black}{0.5}} & \textcolor{black}{$4.0$} & \textcolor{black}{\textemdash{}} & \textcolor{black}{\textemdash{}} & \textcolor{black}{294} & \textcolor{black}{\textemdash{}} & \textcolor{black}{\textemdash{}} & \textcolor{black}{2.3} & \textcolor{black}{0.23}\tabularnewline
 & \textcolor{black}{b} &  &  &  &  &  &  & \textcolor{black}{$4.2$} & \textcolor{black}{$2.6$} & \textcolor{black}{\textemdash{}} & \textcolor{black}{294} & \textcolor{black}{0.12} & \textcolor{black}{no} & \textcolor{black}{27} & \textcolor{black}{2.70}\tabularnewline
 & \textcolor{black}{c} &  &  &  &  &  &  & \textcolor{black}{$4.2$} & \textcolor{black}{$2.6$} & \textcolor{black}{\textemdash{}} & \textcolor{black}{267} & \textcolor{black}{0.12} & \textcolor{black}{yes} & \textcolor{black}{19} & \textcolor{black}{0.35}\tabularnewline
\midrule
\multirow{4}{*}{\textcolor{black}{II}} & \textcolor{black}{a} & \multirow{4}{*}{\textcolor{black}{600}} & \multirow{4}{*}{\textcolor{black}{750}} & \multirow{4}{*}{\textcolor{black}{600}} & \multirow{4}{*}{\textcolor{black}{270}} & \multirow{4}{*}{\textcolor{black}{2400}} & \multirow{4}{*}{\textcolor{black}{0.5}} & \textcolor{black}{$5.4$} & \textcolor{black}{\textemdash{}} & \textcolor{black}{\textemdash{}} & \textcolor{black}{326} & \textcolor{black}{\textemdash{}} & \textcolor{black}{\textemdash{}} & \textcolor{black}{1.8} & \textcolor{black}{0.31}\tabularnewline
 & \textcolor{black}{b} &  &  &  &  &  &  & \textcolor{black}{$5.4$} & \textcolor{black}{$2.5$} & \textcolor{black}{\textemdash{}} & \textcolor{black}{312} & \textcolor{black}{0.11} & yes & \textcolor{black}{6.4} & \textcolor{black}{0.50}\tabularnewline
 & \textcolor{black}{c} &  &  &  &  &  &  & \textcolor{black}{$5.5$} & \textcolor{black}{$4.0$} & \textcolor{black}{\textemdash{}} & \textcolor{black}{375} & \textcolor{black}{0.14} & yes & \textcolor{black}{8.6} & \textcolor{black}{0.36}\tabularnewline
 & \textcolor{black}{d} &  &  &  &  &  &  & \textcolor{black}{$5.5$} & \textcolor{black}{$7.5$} & \textcolor{black}{\textemdash{}} & \textcolor{black}{356} & \textcolor{black}{0.18} & yes & \textcolor{black}{11} & \textcolor{black}{0.55}\tabularnewline
\midrule
\multirow{5}{*}{\textcolor{black}{III}} & \textcolor{black}{a} & \multirow{5}{*}{\textcolor{black}{500}} & \multirow{5}{*}{\textcolor{black}{650}} & \multirow{5}{*}{\textcolor{black}{600}} & \multirow{5}{*}{\textcolor{black}{300}} & \multirow{5}{*}{\textcolor{black}{2400}} & \multirow{5}{*}{\textcolor{black}{1.0 {[}3.0{]}}} & \textcolor{black}{5.3} & \textcolor{black}{\textemdash{}} & \textcolor{black}{\textemdash{}} & \textcolor{black}{647} & \textcolor{black}{\textemdash{}} & \textcolor{black}{\textemdash{}} & \textcolor{black}{1.6} & \textcolor{black}{0.23}\tabularnewline
 & \textcolor{black}{b} &  &  &  &  &  &  & \textcolor{black}{$5.3$} & \textcolor{black}{$1.3$} & \textcolor{black}{\textemdash{}} & \textcolor{black}{667} & \textcolor{black}{0.05} & yes & \textcolor{black}{1.7} & \textcolor{black}{0.19}\tabularnewline
 & \textcolor{black}{c} &  &  &  &  &  &  & \textcolor{black}{$5.5$} & \textcolor{black}{$2.5$} & \textcolor{black}{\textemdash{}} & \textcolor{black}{703} & \textcolor{black}{0.08} & yes & \textcolor{black}{2.0} & \textcolor{black}{0.31}\tabularnewline
 & \textcolor{black}{d} &  &  &  &  &  &  & \textcolor{black}{$5.5$} & \textcolor{black}{$3.7$} & \textcolor{black}{\textemdash{}} & \textcolor{black}{638} & \textcolor{black}{0.10} & yes & \textcolor{black}{2.5} & \textcolor{black}{0.32}\tabularnewline
 & \textcolor{black}{e} &  &  &  &  &  &  & \textcolor{black}{$5.3$} & \textcolor{black}{$2.4$} & 1.9 & \textcolor{black}{664} & \textcolor{black}{0.08} & yes & \textcolor{black}{7.9} & \textcolor{black}{0.27}\tabularnewline
\bottomrule
\end{tabular}\label{growth_details}\centering
\end{table*}
 The oxygen fluxes of 0.5~SCCM (sample series I and II) or 1.0~SCCM
(sample series III), according to Ref.~\textcolor{black}{\onlinecite{Vogt_APL2015}}
correspond to activated oxygen fluxes that can provide growth \textcolor{black}{rates
of $2.2\,\mathrm{\lyxmathsym{\AA}/s}$ and $4.4\,\mathrm{\lyxmathsym{\AA}/s}$,
respectively, under stoichiometric growth conditions. The In cell
(7N purity) temperature was kept at approximately $810\thinspace\mathrm{{^\circ}C}$
(I and II) and $870\thinspace\mathrm{{^\circ}C}$ (III) providing
beam equivalent pressures (BEPs) around $\mathrm{BEP_{\mathrm{In}}=5\times10^{-7}\,\mathrm{mbar}}$,
whereas the Ga cell (7N purity) temperature was varied, so as to provide
samples with different Ga cation concentrations. }One Sn-doped (IGTO)
film, III.e, was grown using an additional Sn flux with \textcolor{black}{$\mathrm{BEP_{\mathrm{Sn}}=1.9\times10^{-8}\,\mathrm{mbar}}$},
as seen in Table~\ref{growth_details}. The metal fluxes limited
the growth rate of the films to approximately $1.3\,\mathrm{\lyxmathsym{\AA}/s}$
(I and II)\textcolor{blue}{{} }\textcolor{black}{to $2.2\,\mathrm{\lyxmathsym{\AA}/s}$}
(III)\textcolor{blue}{,} hence the film development was realized within
the O-rich regime. The substrate was continuously rotated around its
normal axis during the entire growth process at 2 rotations/min in
order to result in films with a homogeneous thickness distribution.

Three separate sample series were synthesized at different combinations
of nucleation and film growth substrate temperatures\textemdash measured
by a thermocouple between the substrate heater and the substrate,
as well as different oxygen fluxes, as indicated in Table~\ref{growth_details}.
Sample series I was realized with the intention to probe the effect
of an approximately 40~nm thick pure In\textsubscript{2}O\textsubscript{3}
buffer layer\textemdash which also acted as a nucleation layer\textemdash between
the substrate and (In\textsubscript{1-\textit{x}}Ga\textsubscript{\textit{x}})\textsubscript{2}O\textsubscript{3}
films, as it was suspected to improve the crystalline quality of the
alloy films. Series II and III were both grown using a buffer layer
(approximately 40~nm and 80~nm thick respectively), based on the
findings examined in the results section III.A. \textcolor{black}{Inspired
by the work of \citet[,][]{Franceschi_PRB2019} the growth was interrupted
after the nucleation of the buffer layer and the substrate was heated
up to $T_{\mathrm{max}}$ as indicated in Table~\ref{growth_details}
to provide films with an enhanced smoothness. The samples of series
(III), in particular, were additionally exposed to an increased O-flux
of 3~SCCM during this interruption. Once $T_{\mathrm{max}}$ was
reached, the sample remained under these conditions for 10~min, before
the substrate was cooled down to the final growth temperature, $T_{\mathrm{growth}}$.}

\textcolor{black}{After the completion of the growth process, the
samples were cooled down to room temperature in vacuum. All heat-up
and cool-down processes were performed under a slow rate of $0.25\,\mathrm{{^\circ}C/s}$
to avoid film delamination. Throughout the entire growth process,
the growth rate was probed by means of laser reflectometry (LR). The
resulting film thicknesses from the total growth time and LR growth
rate match very well those obtained by cross-sectional scanning electron
microscope (SEM) imaging performed later, as specified in Table~\ref{growth_details}.}

\paragraph*{Ex-situ film investigations}

\textcolor{black}{Post-growth energy dispersive X-ray spectroscopy
(EDX) measurements were carried out on the films with a low magnification
so as to identify their average Ga composition. These were performed
using an EDAX }Octane Elect detector\textcolor{black}{{} with a $70\,\mathrm{mm^{2}}$
SDD chip mounted to a Zeiss Ultra55 scanning electron microscope operated
at 7~kV. Standardless quantification using the PhiZAF routine of
the EDAX-Genesis software revealed the percentage of Ga cations that
ultimately got incorporated in the films assuming an oxygen content
corresponding to stoichiometry. The In}\textsubscript{\textcolor{black}{2}}\textcolor{black}{O}\textsubscript{\textcolor{black}{3}}\textcolor{black}{{}
buffer layer could be neglected in the composition analysis because
of the sufficiently large film thickness.}

Out-of-plane X-ray diffraction (XRD) \textit{$2\theta-\omega$} scans
and rocking curves (\textit{\textgreek{w}} scans) were performed within
a PANalytical X'Pert Pro MRD to determine the phase purity and crystalline
quality of the layers. A radiation wavelength $\lambda_{\mathrm{Cu-K\alpha}}=1.5406\,\mathrm{\text{Å}}$\textemdash corresponding
to a photon energy of $8.05\,\mathrm{keV}$\textemdash and a $0.5{^\circ}$
incident-beam slit and $1\,\mathrm{mm}$ receiving slit were used
for the measurements. Apart from these out-of-plane scans, reciprocal
space mapping (RSM) of the (In\textsubscript{1-\textit{x}}Ga\textsubscript{\textit{x}})\textsubscript{2}O\textsubscript{3}
844-reflex in a grazing exit alignment was employed \textcolor{black}{to
gauge the degree of  relaxation of the films and extract their relaxed
lattice parameters. }Moreover, the surface morphology of the films
was characterized by atomic force microscopy in the peak-force tapping
mode of a\textcolor{black}{{} Bruker Dimension Edge AFM system with
ScanAsyst.}

In order to evaluate the single-crystallinity of the films, transmission
electron microscopy (TEM) measurements were performed with a FEI Titan
80 300 operating at 300~kV and equipped with a high annular dark
field detector (HAADF). Additionally, local EDX data was \textcolor{black}{simultaneously}
acquired using a Jeol JEM2200FS TEM at a voltage of 200~kV with an
LN2 free energy dispersive X-ray SD detector to estimate the Ga incorporation
on the nanoscale. The TEM samples were prepared in cross-sectional
view along the $[1\bar{1}0]$ direction of the YSZ substrate by plane-parallel
polishing down to a thickness of 5~\textmu m. For achieving electron
transparency the samples were further thinned by argon ions in an
Gatan PIPS system with an acceleration voltage of 3.5~kV under liquid
nitrogen cooling, followed by a cleaning step where the acceleration
voltage was stepwise reduced to 0.2~kV. 

\textcolor{black}{Raman spectroscopy measurements were performed in
backscattering geometry from the surfaces of the epitaxial films to
investigate the dependence of phonon frequences on the Ga content.
The 325-nm (3.81~eV) line of a He-Cd laser was used for optical excitation.
The incident laser light was focused by a microscope objective onto
the sample surfaces. The backscattered light was collected by the
same objective without analysis of its polarization, then spectrally
dispersed by an 80-cm spectrograph (LabRam HR, Horiba/Jobin Yvon),
and, finally, detected with a liquid-nitrogen-cooled charge-coupled
device (CCD). For the temperature-dependent Raman scattering measurements,
a continuous-flow cryostat (CryoVac) was used in the temperature range
from 10 to 300~K. }

\textcolor{black}{The optical absorption onset of the alloy films
was investigated by spectroscopic ellipsometry in the visible and
ultraviolet spectral range (from 0.5 to 6.5~eV). The measurements
were performed using a Woollam VASE equipped with an autoretarder.
The ellipsometric parameters have been recorded at three angles of
incidence (50°, 60°, and 70°) in order to increase the reliability
of the obtained dielectric function (DF). The DF was analyzed as described
in Ref.~\citep{feneberg_2016}. Both numerical point-by-point fitted
and model dielectric functions are obtained.}

Finally, \textcolor{black}{the electronic states of the bulk of the
films have been investigated by hard x-ray photoelectron spectroscopy
(HAXPES)}. HAXPES measurements were performed at room temperature
at the revolver undulator beamline at BL15XU of SPring-8 using hard\textcolor{black}{{}
x-rays ($hv=5.95\,\mathrm{keV}$).\citep{Ueda_AIPConfProc_2010}}
The corresponding inelastic mean free path (IMFP) of HAXPES for the
In 3\textit{d} core-level photoemission calculated by the Tanuma\textendash Powell\textendash Penn-2M\citep{Tanuma_2006}
is $\lambda=7.29\,\mathrm{nm}$. The probing depth is three times
the IMFP,\citep{POWELL19991} therefore, our HAXPES measurements probe
approximately $22\,\mathrm{nm}$ below the sample surface, which can
reduce the effect of surface Fermi level pinning of In\textsubscript{2}O\textsubscript{3}.
A detailed description of the experimental setup of HAXPES at the
beamline is described elsewhere.\citep{UEDA2013235} A high-resolution
hemispherical electron analyzer (VG Scienta R4000) was used to detect
the photoelectrons. The total energy resolutions of HAXPES was set
to 240~meV. To determine the absolute binding energy, the XPES data
were calibrated against the Au 4\textit{f}\textsubscript{7/2} peak
($84.0\,\mathrm{eV}$) and the Fermi level of Au. Peak fitting of
the XPES data was carried out using the Voigt function after subtracting
the Shirley-type background using the KolXPD software.\citep{kolibrik,Shirley_PhysRevB.5.4709}

\section{Results and discussion}

\subsection{In\protect\textsubscript{2}O\protect\textsubscript{3} buffer layer}

\begin{figure}[h]
\includegraphics[width=8.5cm]{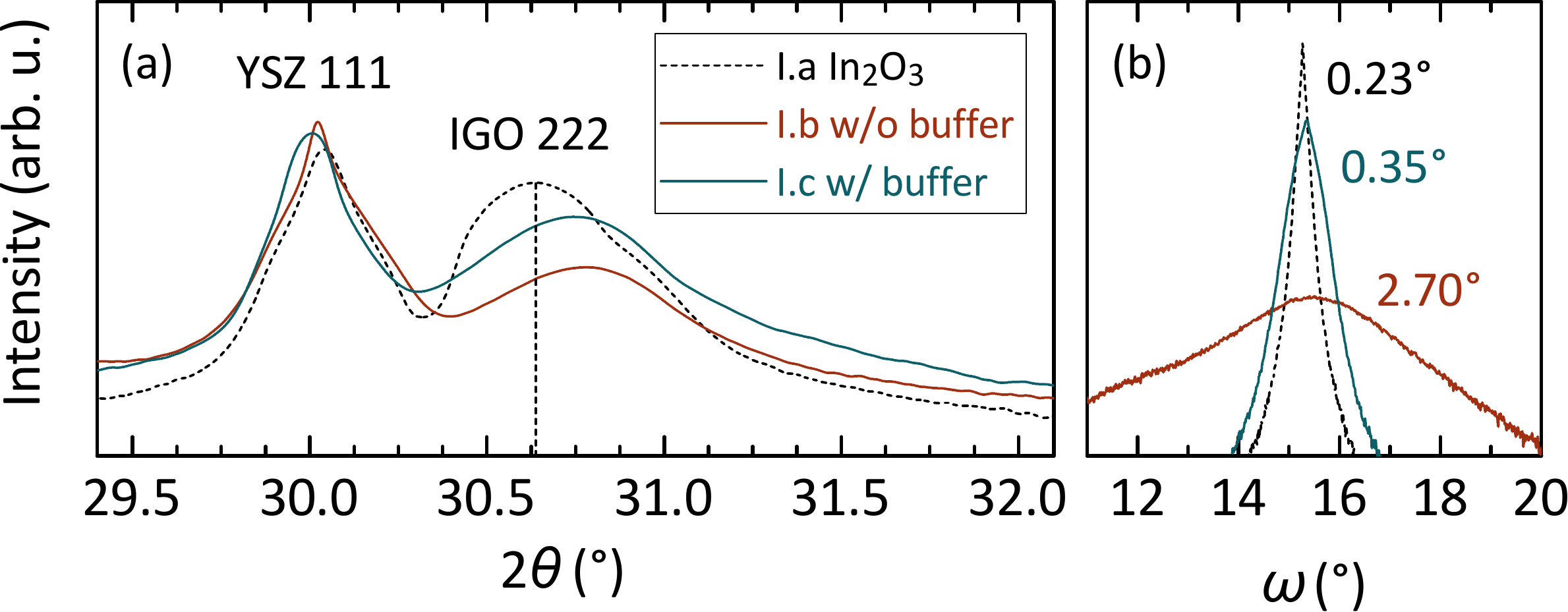}\centering

\caption{(a) X-ray diffraction \textit{$2\theta-\omega$} scan comparing the
two (In\protect\textsubscript{0.88}Ga\protect\textsubscript{0.12})\protect\textsubscript{2}O\protect\textsubscript{3}
samples: with (I.c) and without (I.b) a pure In\protect\textsubscript{2}O\protect\textsubscript{3}
buffer layer between substrate and film (logarithmic vertical scale).
As a reference, the pure In\protect\textsubscript{2}O\protect\textsubscript{3}
film (I.a) is also included as a dotted black line. (b) Rocking curves
of the 222-reflexes shown in the left image. The numbers accompanying
the curves are the corresponding full widths at half maximum (FWHM).
For reference, the bulk YSZ substrates have \textcolor{black}{$\omega_{\mathrm{FWHM}}^{111}=0.02\thinspace{^\circ}$.}
\label{fig:INO_buffer}}
\end{figure}

\textcolor{black}{Despite YSZ being the best-matched substrate for
heteroepitaxy of In}\textsubscript{\textcolor{black}{2}}\textcolor{black}{O}\textsubscript{\textcolor{black}{3}}\textcolor{black}{,
growth on it poses certain challenges. The growth is realized under
the Volmer\textendash Weber growth mode, where the film nucleates
into separate islands, rather than wet the substrate.\citep{Bierwagen_JAP_2010}
To create more favorable wetting conditions and force nucleation,
the growth temperature of the substrate needs to be reduced for the
growth of the initial few nm of the layer, which can have a negative
impact on film smoothness and overall quality. Besides this, the resulting
film will have a tensile mismatch of 1.6\,\% to the substrate.}

\textcolor{black}{The incorporation of Ga into the In}\textsubscript{\textcolor{black}{2}}\textcolor{black}{O}\textsubscript{\textcolor{black}{3}}\textcolor{black}{{}
lattice is expected to further aggravate the challenges of growth
on YSZ.} For example, In\textsubscript{2}O\textsubscript{3} has
a lattice constant of 10.117~Å\citep{Marezio1966}\textemdash slightly
smaller than twice that of the substrate YSZ, at 10.28~Å, whereas
cubic Ga\textsubscript{2}O\textsubscript{3} has a theoretically
predicted lattice constant of approximately $9.190-9.401\,\mathrm{\mathring{A}}$.\citep{peelaers2015,Yoshioka_2007}
A lattice mismatch of 1.6\,\% could already lead to the formation
of lattice defects, such as dislocations or grain boundaries, to compensate
for the change in lattice parameter, and hence result in films with
low crystalline quality.\textcolor{black}{{} Particularly in alloy films,
this effect would be exacerbated. For example, an (In}\textsubscript{\textcolor{black}{1-}\textit{\textcolor{black}{x}}}\textcolor{black}{Ga}\textsubscript{\textcolor{black}{x}}\textcolor{black}{)}\textsubscript{\textcolor{black}{2}}\textcolor{black}{O}\textsubscript{\textcolor{black}{3}}\textcolor{black}{{}
film with $x=0.10$ would have a lattice mismatch of up to 2.5\,\%
to the YSZ substrate\textemdash assuming that the lattice parameter
of the film decreases with }\textit{\textcolor{black}{x}}\textcolor{black}{{}
according to Vegard's law (see next section for further explanation).
This increased lattice mismatch between layer and substrate can result
in additional crystal defects.}\textcolor{blue}{{} }One approach to
circumvent this and partially release the strain, would be to not
directly grow the (In,Ga)\textsubscript{2}O\textsubscript{3} layers
on top of YSZ, but rather aim for an initial nucleation of a pure
In\textsubscript{2}O\textsubscript{3} thin layer on YSZ, and then
proceed with the growth of the (In,Ga)\textsubscript{2}O\textsubscript{3}
on top of it. This In\textsubscript{2}O\textsubscript{3} buffer
layer will then effectively act as a secondary substrate with a matching
crystal structure to the alloy, which should aid film orientation,
and, additionally, improve the nucleation conditions. Furthermore,
the buffer layer allows for higher growth at temperatures and could,
thus, result in a smoother film. \textcolor{black}{This technique
should also enhance the solubility at the low-}\textit{\textcolor{black}{x}}\textcolor{black}{{}
end, according to the calculations of \citet{Maccioni_2016} on the
miscibility of the alloy, where the miscibility of (In}\textsubscript{\textcolor{black}{1-}\textit{\textcolor{black}{x}}}\textcolor{black}{Ga}\textsubscript{\textit{\textcolor{black}{x}}}\textcolor{black}{)}\textsubscript{\textcolor{black}{2}}\textcolor{black}{O}\textsubscript{\textcolor{black}{3}}\textcolor{black}{{}
is enhanced in epitaxial growth upon In}\textsubscript{\textcolor{black}{2}}\textcolor{black}{O}\textsubscript{\textcolor{black}{3}}\textcolor{black}{,
compared to the free-standing alloy.}

\begin{figure*}[t]
\includegraphics[width=16cm]{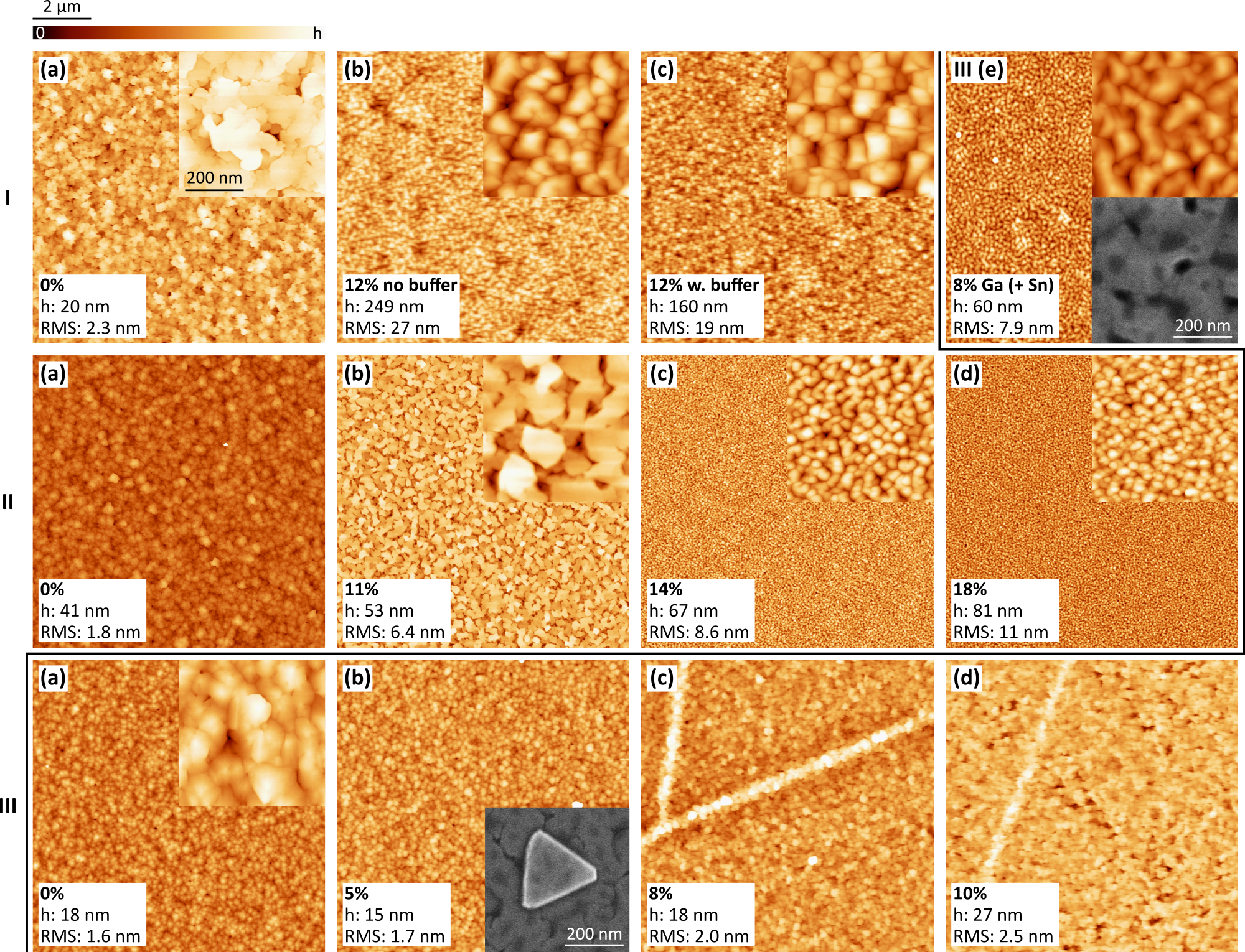}\centering

\caption{10\texttimes 10~\textgreek{m}m\protect\textsuperscript{2} atomic
force micrographs of the (In\protect\textsubscript{1-$x$}Ga\protect\textsubscript{$x$})\protect\textsubscript{2}O\protect\textsubscript{3}
films with growth parameters as described in detail in Table~\ref{growth_details}
(insets: 1\texttimes 1~\textgreek{m}m\protect\textsuperscript{2}).
Nucleation and growth temperatures are highest for sample series I
and lowest for series III and Ga content $x$, increases from left
to right, explicitly indicated in percentage form. The heightscale
of the images is indicated by the colored bar at the top and corresponding
$h$ in each image. The RMS value\textcolor{black}{, also mentioned
in Table~\ref{growth_details},} indicates the roughness of the films.
The grayscale insets in III.b and III.e are close-up scanning electron
microscopy (SEM) images showing the triangular shape of features on
the film surface.\label{fig:AFM_all}}
\end{figure*}

A comparison of the crystalline qualities of the films with and without
a buffer layer is presented in Figure~\ref{fig:INO_buffer} based
on symmetric XRD $2\theta-\omega$ scans and corresponding rocking
curves. The shift of the \textcolor{black}{(In}\textsubscript{\textcolor{black}{1-}\textit{\textcolor{black}{x}}}\textcolor{black}{Ga}\textsubscript{\textit{\textcolor{black}{x}}}\textcolor{black}{)}\textsubscript{\textcolor{black}{2}}\textcolor{black}{O}\textsubscript{\textcolor{black}{3}}
layer reflex towards larger angles compared to the position of the
In\textsubscript{2}O\textsubscript{3} layer reflex indicates a decrease
in the lattice parameter, as expected. Evidently, a pure In\textsubscript{2}O\textsubscript{3}
buffer layer on top of YSZ results in an \textcolor{black}{(In}\textsubscript{\textcolor{black}{1-}\textit{\textcolor{black}{x}}}\textcolor{black}{Ga}\textsubscript{\textit{\textcolor{black}{x}}}\textcolor{black}{)}\textsubscript{\textcolor{black}{2}}\textcolor{black}{O}\textsubscript{\textcolor{black}{3}}
film with a significantly better quality compared to the one grown
on top of a Ga-containing nucleation layer. This is further confirmed
by the AFM images in Fig.~\ref{fig:AFM_all}, which shows that the
sample grown on a buffer layer (I.c) is significantly smoother and
has better defined structures than the one without the buffer (I.b).
Based on this finding, all further samples grown for this study included
an In\textsubscript{2}O\textsubscript{3} buffer for improved surface
smoothness and crystallinity. 

\subsection{Structure dependence on Ga content}

\subsubsection*{Surface morphology}

As In\textsubscript{2}O\textsubscript{3} does not easily wet the
YSZ substrate,\citep{Bierwagen_JAP_2010} the substrate temperature
used for the growth of\textemdash especially\textemdash the initial
few nm of the epilayer has a clear effect on its crystalline quality,
which is also confirmed by the comparison of three pure In\textsubscript{2}O\textsubscript{3}
films (I.a, II.a, III.a) grown at different temperatures, in the AFM
images of Fig.~\ref{fig:AFM_all}. An overall increasing roughness
and density of morphological features with increasing Ga content can
be observed.\textcolor{black}{{} The average grain size decreases with
Ga, in accordance with the findings of \citet[.][]{kong_2010}}\textcolor{blue}{{}
}This indicates that the incorporation of Ga leads to a lower surface
diffusion length. The growth temperature seemingly also affects the
surface diffusion length, with lower temperatures resulting in smoother
films with smaller features. However, this result is probably due
to the lower nucleation temperature, which enhances wetting.

\subsubsection*{Phase purity, lattice parameter, and compositional (in)homogeneity}

\begin{figure}[h]
\includegraphics[width=6cm]{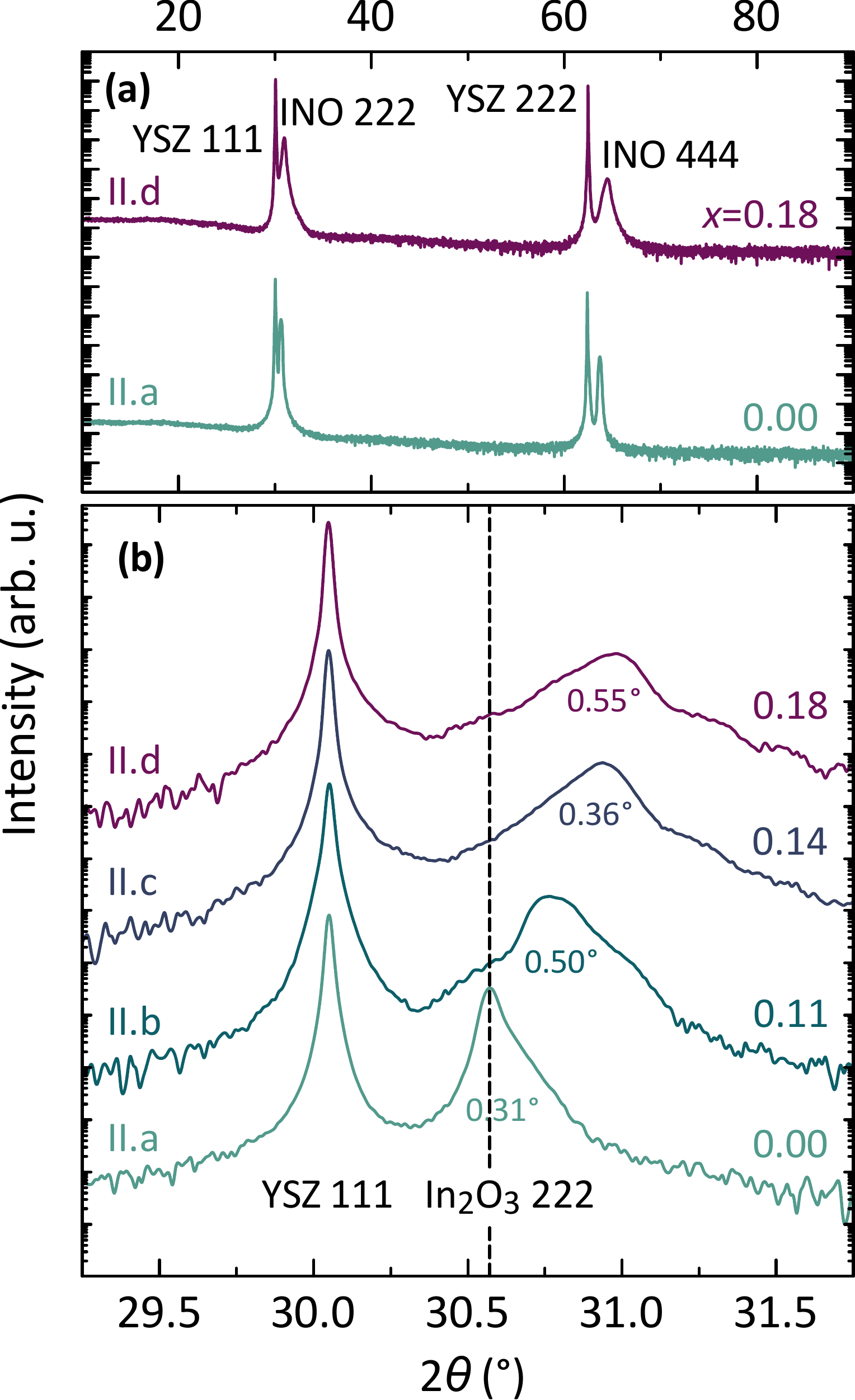}\centering

\caption{Symmetric $2\theta-\omega$ XRD scans of sample series (II). (a) Broad
scan including two reflex orders shows no additional crystalline phases.
(b) Narrow scan around the 111-reflex of the substrate (left peak)
and the 222-reflex of the layer (right). The layer peak shifts towards
larger angles with additional Ga. The numbers underneath the layer
peaks are the FWHM of the rocking curves of the respective layer peak.
\label{fig:XRD_out-of-plane}}
\end{figure}

\textcolor{black}{As seen in the wide range symmetric $2\theta-\omega$
scans XRD scans in Fig.~\ref{fig:XRD_out-of-plane}~(a), the film
with the highest Ga content, (In}\textsubscript{\textcolor{black}{0.82}}\textcolor{black}{Ga}\textsubscript{\textcolor{black}{0.18}}\textcolor{black}{)}\textsubscript{\textcolor{black}{2}}\textcolor{black}{O}\textsubscript{\textcolor{black}{3}}\textcolor{black}{,
exhibits only one pure, 111-oriented cubic phase, in contrast to earlier
works on In}\textsubscript{\textcolor{black}{2}}\textcolor{black}{O}\textsubscript{\textcolor{black}{3}}\textcolor{black}{{}
ceramics\citep{regoutz_2015} and nanowires\citep{Chun_APL2004} with
comparable or even lower Ga contents, where several orientations are
reported. Moreover, no secondary crystalline Ga}\textsubscript{\textcolor{black}{2}}\textcolor{black}{O}\textsubscript{\textcolor{black}{3}}\textcolor{black}{{}
phases are observed, in contrast to a previous investigation on films
grown by PA-MBE containing higher amounts of Ga.\citep{Vogt_PhD}
The observed low full width at half maximum (FWHM) values from the
rocking curves in Fig.~\ref{fig:XRD_out-of-plane}~(b), $\omega_{\mathrm{FWHM}}^{222}$,
indicate high crystalline quality. However, an overall increase of
the $\omega_{\mathrm{FWHM}}^{222}$ with increasing Ga content is
exhibited, which would be consistent with slight deterioration of
film quality. The lower intensity features on the right of the (In}\textsubscript{\textcolor{black}{1-}\textit{\textcolor{black}{x}}}\textcolor{black}{Ga}\textsubscript{\textit{\textcolor{black}{x}}}\textcolor{black}{)}\textsubscript{\textcolor{black}{2}}\textcolor{black}{O}\textsubscript{\textcolor{black}{3}}\textcolor{black}{{}
peak could point towards compositional inhomogeneity in the film,
however, considering the $2\theta$ values at which they arise, they
could also be attributed to interference fringes from the high-quality
thin In}\textsubscript{\textcolor{black}{2}}\textcolor{black}{O}\textsubscript{\textcolor{black}{3}}\textcolor{black}{{}
buffer layer at the interface.}

\textcolor{black}{}
\begin{figure}[h]
\textcolor{black}{\includegraphics[width=8cm]{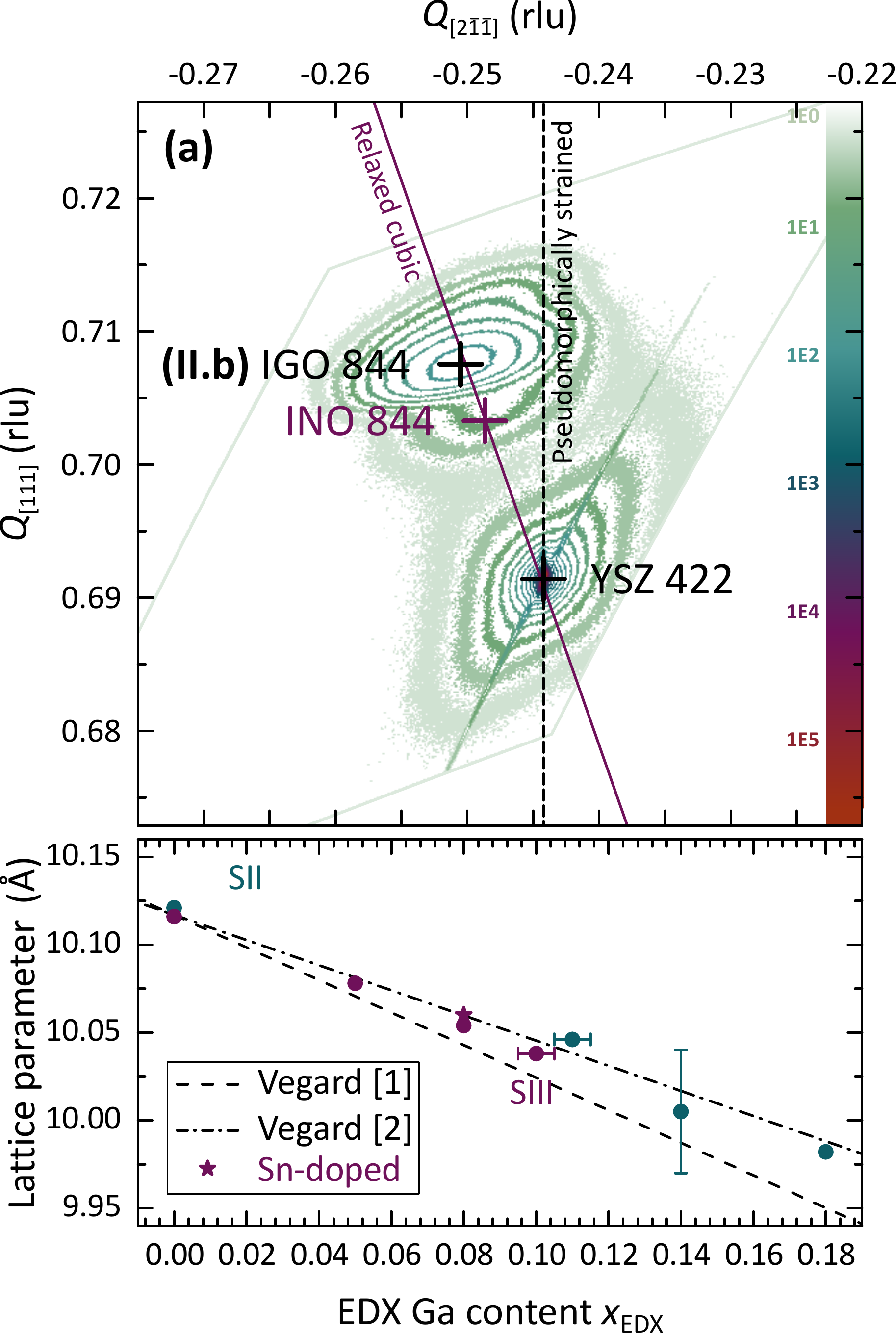}\centering}

\textcolor{black}{\caption{(a) Example of an RSM at an asymmetrical grazing exit alignment, showing
the 422-reflex of the substrate and 844-reflex of the layer (performed
on sample II.b). The peak positions of t\textcolor{black}{he YSZ substrate
and the (In}\protect\textsubscript{\textcolor{black}{0.89}}\textcolor{black}{Ga}\protect\textsubscript{\textcolor{black}{0.11}}\textcolor{black}{)}\protect\textsubscript{\textcolor{black}{2}}\textcolor{black}{O}\protect\textsubscript{\textcolor{black}{3}}\textcolor{black}{film
(II.b) are marked by crosses and labeled accordingly, along with the
theoretical position of pure In}\protect\textsubscript{\textcolor{black}{2}}\textcolor{black}{O}\protect\textsubscript{\textcolor{black}{3}}\textcolor{black}{{}
(purple). $Q_{\mathrm{[2\bar{1}\bar{1]}}}$ and $Q_{\mathrm{[111]}}$
expressed in reciprocal lattice units (rlu) correspond to the in-
and out-of-plane directions respectively. The maps are drawn by equi-intensity
lines using a color coded, logarithmic scale as specified. The vertical
dashed and oblique solid lines are guides for pseudomorphically strained
and fully relaxed layers, respectively. (b) Ga cation content measured
by EDX, $x_{\mathrm{EDX}}$ and corresponding relaxed lattice parameters,
$a_{0}$, extracted from the RSM measurements. The dashed lines correspond
to Vegard's law calculated for the lattice constants of cubic bixbyite
Ga}\protect\textsubscript{\textcolor{black}{2}}\textcolor{black}{O}\protect\textsubscript{\textcolor{black}{3}}\textcolor{black}{in
{[}1{]} Ref.~\citet{peelaers2015} and {[}2{]} Ref.~\citet{Yoshioka_2007}.
The horizontal error bar for samples SII.b and SIII.d is representative
of the 1\,\% uncertainty in the $x_{\mathrm{EDX}}$ and applies to
all samples.\label{fig:XRD_RSM}}}
}
\end{figure}

\textcolor{black}{The observed shift of the layer peak towards larger
$2\theta$ values is expected, as it reflects the decrease of lattice
constant, from that of In}\textsubscript{\textcolor{black}{2}}\textcolor{black}{O}\textsubscript{\textcolor{black}{3}}\textcolor{black}{{}
towards the theoretically predicted one of cubic Ga}\textsubscript{\textcolor{black}{2}}\textcolor{black}{O}\textsubscript{\textcolor{black}{3}}\textcolor{black}{.
Since both parent components, i.e., In}\textsubscript{\textcolor{black}{2}}\textcolor{black}{O}\textsubscript{\textcolor{black}{3}}\textcolor{black}{{}
and the theoretically predicted bixbyite Ga}\textsubscript{\textcolor{black}{2}}\textcolor{black}{O}\textsubscript{\textcolor{black}{3}}\textcolor{black}{,
have the same crystal structure in their pure form, Vegard's law can
be applied (Eq.~\ref{eq:Vegard}) to obtain the relaxed lattice parameter
of the alloy.}

\textcolor{black}{
\begin{equation}
a_{\mathrm{0,(In_{1-\textit{x}}Ga_{\textit{x}})_{2}O_{3}}}=(1-x)a_{\mathrm{0,In_{2}O_{3}}}+xa_{\mathrm{0,Ga_{2}O_{3}}}\label{eq:Vegard}
\end{equation}
}

\textcolor{black}{However, the peak position of measurements like
those in Fig.~\ref{fig:XRD_out-of-plane} provide information only
about the out-of-plane lattice parameter. In order to gain knowledge
of the relaxed lattice parameter of the layers\textemdash which is
affected by both in- and out-of-plane strain, RSMs have been obtained
for the films of series II and III. These were performed at an asymmetrical
grazing exit alignment at a range of $2\theta$ and $\omega$ values
that included the 4\,2\,2-reflex of the YSZ substrate and 8\,4\,4-reflex
of the (In}\textsubscript{\textcolor{black}{1-}\textsl{\textcolor{black}{x}}}\textcolor{black}{Ga}\textsubscript{\textsl{\textcolor{black}{x}}}\textcolor{black}{)}\textsubscript{\textcolor{black}{2}}\textcolor{black}{O}\textsubscript{\textcolor{black}{3}}\textcolor{black}{{}
film. An example of such a map can be seen in Fig.~\ref{fig:XRD_RSM}
(a). The (4\,2\,2) and (8\,4\,4) planes can be decomposed into
a component parallel and perpendicular to the lattice plane as}

\begin{subequations}
\begin{equation}
(422)=\frac{2}{3}(2\bar{1}\bar{1})+\frac{8}{3}(111)\label{eq:plane_decomposition-422}
\end{equation}

\begin{equation}
(844)=\frac{4}{3}(2\bar{1}\bar{1})+\frac{16}{3}(111)\label{eq:plane_decomposition_844}
\end{equation}
\end{subequations}

\textcolor{black}{The corresponding reciprocal lattice vectors $Q_{\mathrm{[2\bar{1}\bar{1}]}}$
and $Q_{\mathrm{[111]}}$ along those two directions are}

\begin{equation}
Q_{\mathrm{[2\bar{1}\bar{1}]}}=\frac{\cos(\omega)-\cos(2\theta-\omega)}{2}\label{eq:reciprocal_lattice_vectors_in}
\end{equation}
\begin{equation}
Q_{\mathrm{[111]}}=\frac{\sin(\omega)+\sin(2\theta-\omega)}{2}\label{eq:reciprocal_lattice_vectors_out}
\end{equation}

where $\omega$ and $2\theta$ are the incidence and diffraction angles
in radians, respectively.\textcolor{black}{{} All cubic materials should
have the same ratio of in- and out-of-plane lattice components, i.e.,
for a fully relaxed layer, in map plotted in terms of $Q_{\mathrm{[2\bar{1}\bar{1}]}}$
and $Q_{[111]}$ such as the one in Fig.~\ref{fig:XRD_RSM} (a),
we would expect the (In}\textsubscript{\textcolor{black}{1-}\textsl{\textcolor{black}{x}}}\textcolor{black}{Ga}\textsubscript{\textsl{\textcolor{black}{x}}}\textcolor{black}{)}\textsubscript{\textcolor{black}{2}}\textcolor{black}{O}\textsubscript{\textcolor{black}{3}}\textcolor{black}{{}
844 peak to lie on the same line connecting the $(Q_{\mathrm{[2\bar{1}\bar{1}]}},Q_{\mathrm{[111]}})=(0,0)$
point and the 422 peak of the cubic YSZ substrate. For a pseudomorphically
strained film, the grown film would assume the in-plane lattice parameter
of the substrate, hence the in-plane component $Q_{\mathrm{[2\bar{1}\bar{1}]}}$
of both should be equal. Based on this, it can be concluded that the
sample II.b is almost fully relaxed. The same applies to all other
films grown for this study, the maps of which are not plotted here.}

\textcolor{black}{The in- and out-of-plane lattice spacing between
adjacent (In}\textsubscript{\textcolor{black}{1-}\textit{\textcolor{black}{x}}}\textcolor{black}{Ga}\textsubscript{\textit{\textcolor{black}{x}}}\textcolor{black}{)}\textsubscript{\textcolor{black}{2}}\textcolor{black}{O}\textsubscript{\textcolor{black}{3}}\textcolor{black}{{}
lattice planes can be calculated from the $Q_{\mathrm{[2\bar{1}\bar{1}]}}$
and $Q_{\mathrm{\mathrm{[111]}}}$ as}

\begin{subequations}
\textcolor{black}{
\begin{equation}
d_{\mathrm{(\frac{8}{3}\frac{\bar{4}}{3}\frac{\bar{4}}{3})}}=\frac{\lambda}{2\lvert Q_{\mathrm{\mathrm{[2\bar{1}\bar{1}]}}}\lvert}\label{eq:plane_spacing_in}
\end{equation}
 }
\begin{equation}
d_{\mathrm{(\frac{16}{3}\frac{16}{3}\frac{16}{3})}}=\frac{\lambda}{2Q_{\mathrm{\mathrm{[111]}}}}\label{eq:plane_spacing_out}
\end{equation}
\end{subequations}

where $\lambda$ is the wavelength of the x-rays.\textcolor{black}{{}
The corresponding in- and out-of-plane (In}\textsubscript{\textcolor{black}{1-}\textit{\textcolor{black}{x}}}\textcolor{black}{Ga}\textsubscript{\textit{\textcolor{black}{x}}}\textcolor{black}{)}\textsubscript{\textcolor{black}{2}}\textcolor{black}{O}\textsubscript{\textcolor{black}{3}}\textcolor{black}{{}
lattice constants are calculated using the in- and out-of-plane components
of the (844) plane given by Eq.~\ref{eq:plane_decomposition_844}
and}

\begin{subequations}
\textcolor{black}{
\begin{equation}
a_{\mathrm{[2\bar{1}\bar{1}]}}=\frac{4\sqrt{6}}{3}d_{\mathrm{(\frac{8}{3}\frac{\bar{4}}{3}\frac{\bar{4}}{3})}}\label{eq:lattice_constant_in}
\end{equation}
}

\begin{equation}
a_{\mathrm{[111]}}=\frac{16\sqrt{3}}{3}d_{\mathrm{(\frac{16}{3}\frac{16}{3}\frac{16}{3})}}\label{eq:lattice_constant_out}
\end{equation}
\end{subequations}

\textcolor{black}{accordingly. Based on the values for $a_{\mathrm{[2\bar{1}\bar{1}]}}$
and $a_{\mathrm{[111]}}$ extracted from the maps and the definition
of the Poisson ratio\citep{Birkholz_2006} (Eq.~\ref{eq:Poisson},
where $\varepsilon_{\mathrm{[2\bar{1}\bar{1}],[111]}}$ are the in-
and out-of-plane strain respectively), one can calculate the total
relaxed lattice parameter, $a_{\mathrm{0}}$ , whilst taking into
account that in the case of a cubic crystal the in- and out-of-plane
relaxed lattice parameters should be equal.
\begin{equation}
\frac{\varepsilon_{\mathrm{[2\bar{1}\bar{1}]}}}{\varepsilon_{\mathrm{[111]}}}=\frac{a_{\mathrm{[2\bar{1}\bar{1}]}}-a_{\mathrm{0}}}{a_{\mathrm{[111]}}-a_{0}}=-\frac{2\nu}{1-\nu}\label{eq:Poisson}
\end{equation}
}

\textcolor{black}{Here we used $\nu=0.31$ obtained experimentally
by \citet{Zhang_PhysRevB.84.233301} for 111-oriented In}\textsubscript{\textcolor{black}{2}}\textcolor{black}{O}\textsubscript{\textcolor{black}{3}}\textcolor{black}{{}
grown on YSZ. Fig.~\ref{fig:XRD_RSM}~(b) shows the relaxed lattice
parameters, $a_{\mathrm{0}}$, of the alloy films extracted from the
RSM measurements as a function of Ga content, as measured by EDX.
The obtained $a_{\mathrm{0}}$ agree reasonably well with the experimental
results documented in the earlier work of \citet{regoutz_2015} on
(In,Ga)}\textsubscript{\textcolor{black}{2}}\textcolor{black}{O}\textsubscript{\textcolor{black}{3}}\textcolor{black}{{}
ceramics. The plot reveals a roughly linear dependence of the lattice
constant on the alloy composition, hence, our films largely follow
Vegard's law within the margins of error. The uncertainty in $x$
is approximately 1\,\% for all samples and uncertainty in $a_{\mathrm{0}}$
of sample II.c as depicted in Fig.~\ref{fig:XRD_RSM}~(b) is due
to significant discrepancies among the various obtained RSMs. The
limits posed by the two dashed lines have been calculated assuming
Vegard's law for the lattice constants of cubic bixbyite Ga}\textsubscript{\textcolor{black}{2}}\textcolor{black}{O}\textsubscript{\textcolor{black}{3}}\textcolor{black}{{}
reported in Refs.~\citet{peelaers2015} (9.190~Å, extracted from
Fig.~3 therein) and \citet{Yoshioka_2007} (9.40~Å). The agreement
with Vegard's law implies the incorporation of Ga cations on In sites
and does not indicate a solubility limit within the $x$-range studied
here, that is up to $x=0.18$. A linear fit of the lattice constant
data in Fig.~\ref{fig:XRD_RSM}~(b) lets us estimate an experimental
value for the hypothetical cubic bixbyite phase of Ga}\textsubscript{\textcolor{black}{2}}\textcolor{black}{O}\textsubscript{\textcolor{black}{3}}\textcolor{black}{{}
at $9.365\thinspace(\pm0.018)\,\text{Å}$.}

\textcolor{black}{Unfortunately, the low intensity and large width
of the layer reflexes in the RSMs did not allow the extraction of
separate and precise in- and out-of-plane lattice parameters\textemdash and
therefore strains\textemdash that can be systematically explained
for the samples with $x\geq0.10$. This is because the relaxed lattice
parameter, $a_{0}$, is relatively less affected by the reflex position,
as compared to the strain. However, comparing their In}\textsubscript{\textcolor{black}{2}}\textcolor{black}{O}\textsubscript{\textcolor{black}{3}}\textcolor{black}{{}
reference samples (II.a, III.a), series II seems to be less strained
than series III, which was grown at lower nucleation and final substrate
temperatures, with II.a showing an in-plane strain of $\varepsilon_{\mathrm{[111]}}^{\mathrm{II.a}}\simeq0.07\%$
and sample III.a $\varepsilon_{\mathrm{[111]}}^{\mathrm{III.a}}\simeq0.15\thinspace\%$.
We do not expect the alloy films to be significantly more strained
and presume an upper strain limit of approximately $0.2\thinspace\%$.}

\begin{figure*}
\includegraphics[width=16cm]{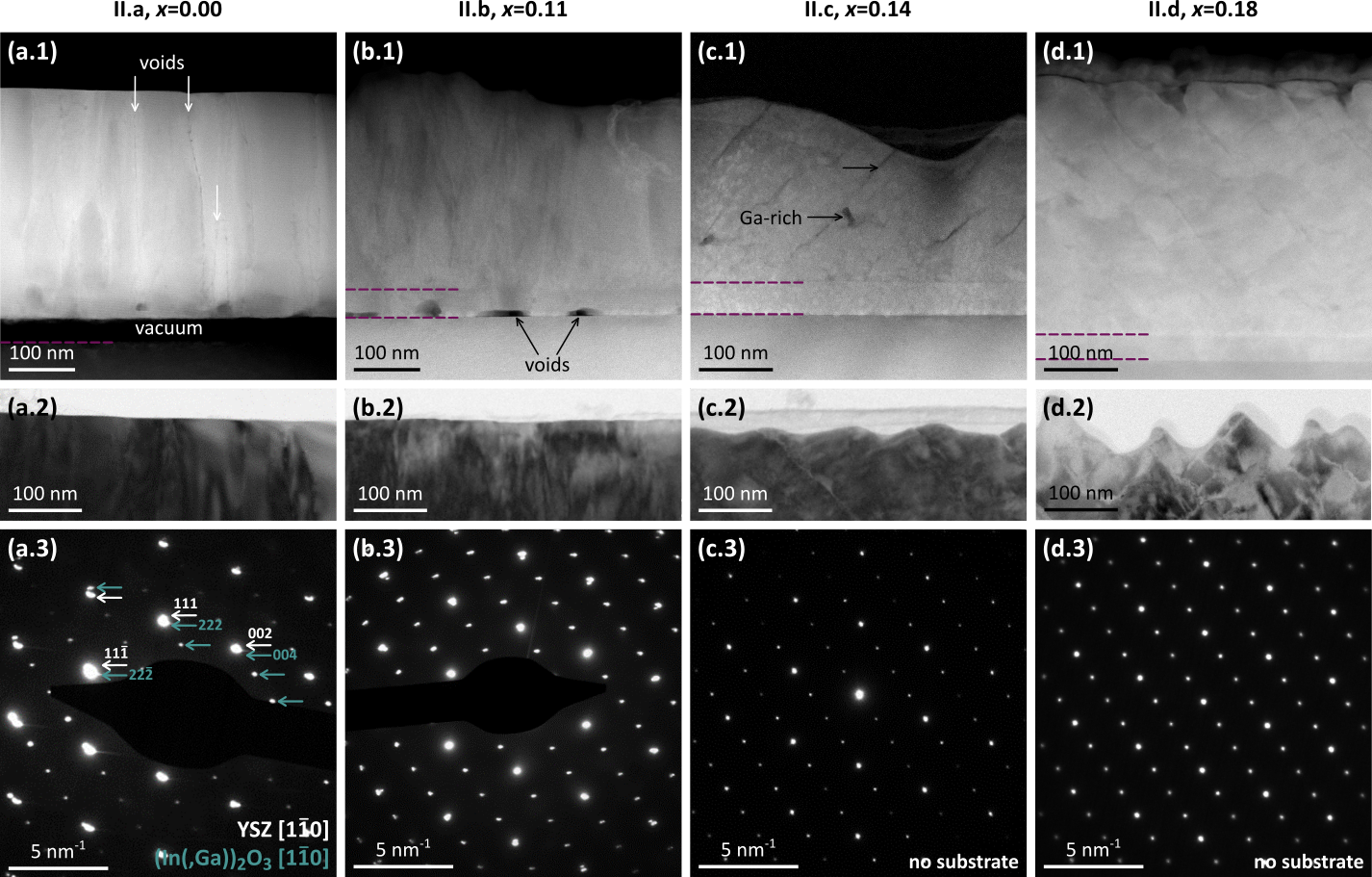}\centering

\caption{\textcolor{black}{Top (a-d.1): HAADF-STEM cross-section images of
(In}\protect\textsubscript{\textcolor{black}{1-}\textit{\textcolor{black}{x}}}\textcolor{black}{Ga}\protect\textsubscript{\textit{\textcolor{black}{x}}}\textcolor{black}{)}\protect\textsubscript{\textcolor{black}{2}}\textcolor{black}{O}\protect\textsubscript{\textcolor{black}{3}}\textcolor{black}{{}
films with }\textit{\textcolor{black}{x}}\textcolor{black}{{} ranging
between 0.00 and 0.18, as indicated. The dashed lines indicate the
interfaces between substrate, buffer layer, and film.}\textcolor{red}{{}
}\textcolor{black}{Middle (a-d.2): Bright field TEM images of the
real growth surfaces. Bottom (a-d.3): Electron diffraction patterns
of each sample taken in the $[11\bar{0}]$ orientation, with spots
belonging to the YSZ substrate and the layer indexed in white and
teal respectively.}\label{fig:TEM_1}}
\end{figure*}

\begin{figure}[h]
\includegraphics[width=6cm]{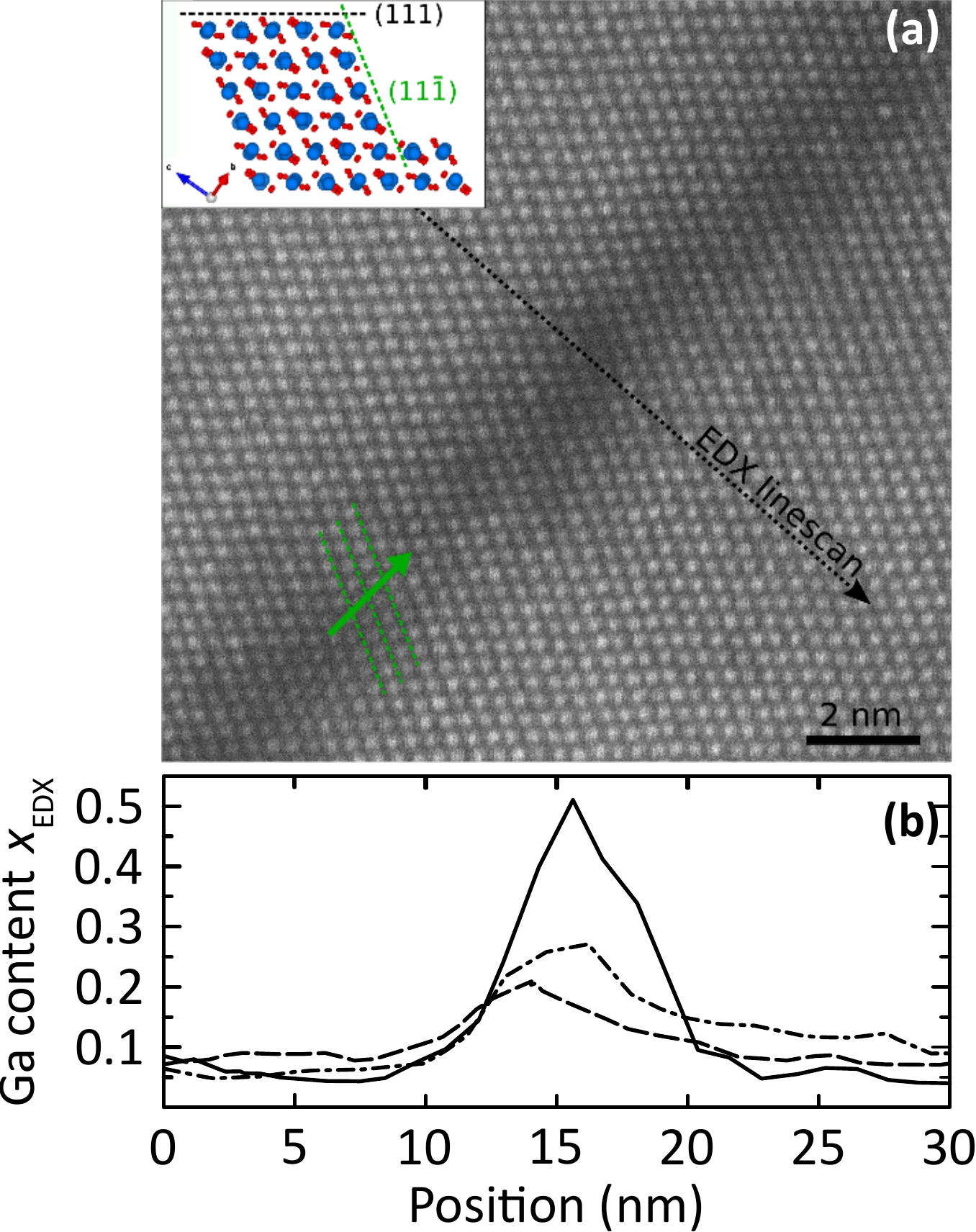}\centering

\caption{(a) HAADF-STEM high magnification image capturing one of the dark
stripes traveling through the (In\protect\textsubscript{0.86}Ga\protect\textsubscript{0.14})\protect\textsubscript{2}O\protect\textsubscript{3}
(II.c). \textcolor{black}{Bright spots correspond to projected atomic
(In,Ga) columns (oxygen is too light to produce visible contrast),
and the observed pattern fits to the bixbyite model structure. }The
atomic pattern is resolved and fits to the model structure of cubic
bixbyite phase in $[11\bar{0}]$ orientation. (b) EDX line scans across
three different dark stripes showing an increased Ga incorporation
in these areas.\label{fig:TEM_2}}
\end{figure}

The films of sample series II have\textcolor{black}{{} been investigated
by TEM to examine their single-crystallinity and overall film quality.
At the top of Fig.~\ref{fig:TEM_1}~(a.1-d.1), high angle annular
dark field scanning transmission electron microscopy (HAADF-STEM)
images, in which the contrast is proportional to the mean atomic number,
of the four epitaxial layers are compared. The surfaces there are
affected by the ion milling process, in which the sample is thinned
down to electron transparency. The real growth surfaces are shown
right underneath (Fig.~\ref{fig:TEM_1}~a.2-d.2) as bright field
TEM images. Some voids can be observed traveling in the form of thin
vertical lines through the layer for the pure In}\textsubscript{\textcolor{black}{2}}\textcolor{black}{O}\textsubscript{\textcolor{black}{3}}\textcolor{black}{{}
as indicated by white arrows, and as more extended regions in the
film with $x=0.11$. The films with }\textit{\textcolor{black}{x}}\textcolor{black}{=0.14
and }\textit{\textcolor{black}{x}}\textcolor{black}{=0.18 do not show
such strong voids at the interface and surface and appear rougher
with peak to valley distances of a few tens of a nm. This suggests
that the addition of Ga induces a different growth mode along facets
due to a destabilization of the (111) plane. Another distinct feature
in samples II.c and II.d with the nominally highest Ga contents are
the dark stripes traveling diagonally upward at a fixed angle through
the layer. These features can be identified as regions of increased
Ga incorporation according to the EDX analysis of sample II.c, as
shown below (Fig.~\ref{fig:TEM_2}). The combined electron diffraction
patterns of substrate and film, in Fig.~\ref{fig:TEM_1} (a.3-d.3),
show the presence of a single cubic bixbyite crystalline phase in
all samples. All points can be linked to either those of the cubic
structure of YSZ or those of the cubic bixbyite structure of the film.
Since these lattices have similar cubic symmetry with an almost identical
lattice parameter\textemdash considering a doubling of the periodicity
for the bixbyite phase\textemdash some spots of the distinct phases
are overlapping and have been indexed twice.}

\textcolor{black}{EDX point scans on sample II.b with $x=0.11$ performed
in darker and brighter intensity regions of the film show Ga content
fluctuations of only 1\,\% (not shown here). This indicates a homogeneous
incorporation of the Ga cations in the cubic bixbyite lattice. However,}\textit{\textcolor{black}{{}
}}\textcolor{black}{sample II.c with $x=0.14$ exhibits areas with
increased Ga contents varying from $\,x=0.20$ up to 0.50 compared
to the surrounding ``matrix'', which features an average Ga incorporation
of approximately $x=0.08$. This has been identified by EDX line scans
crossing the dark features, as shown in Fig.~\ref{fig:TEM_2} (b).
One such dark stripe is imaged by HAADF-STEM at high magnification
in Fig.~\ref{fig:TEM_2} (a) in the $[1\bar{1}0]$ zone axis orientation
of the lattice. The bixbyite structure is preserved throughout the
dark stripe without the formation of an additional phase or any lattice
defects, which confirms the electron diffraction data that the layer
is single-phase. The EDX line scans also show a dip in O content coinciding
with the Ga increase (cf. supplement,\citealp{SUPPLEMENT} Fig.~2
therein), indicating void formation as well. The rough surface, the
void formation, and the defined orientation of the Ga-rich stripes
suggests a faceted 3D island growth, with phase separation resulting
from the preferential incorporation of Ga at edges of voids and on
well-defined facets. The edge of preferential incorporation seems
to correspond to the $(11\bar{1})$ plane (equivalent to 111), as
indicated in Fig.~\ref{fig:TEM_2} (a), and the edge travels both
laterally and vertically (in the direction of the green arrow) through
the layer. Similar Ga-rich features are observed in the MBE growth
of AlGaN, where enhanced Ga incorporation is observed on step edges
due to a higher Ga desorption rate on the terraces.\citep{Mayboroda_JAP_2017}
The $x=0.18$ sample (not shown here) exhibits similar darker stripes
and defects due to relaxation processes through misfit dislocations,
as well as grains and initial signs of phase separation and amorphicity,
but maintains a single cubic phase throughout.}

\textcolor{black}{Hence, despite the preservation of the bixbyite
phase and the good agreement with Vegard's law, Ga is inhomogeneously
distributed in the samples with $x>0.11$.}

\subsubsection*{\textcolor{black}{Saturation of Raman phonon mode shift, (optical)
absorption edge, and Ga 2p core level position}}

\textcolor{black}{}
\begin{figure}[h]
\textcolor{black}{\includegraphics[width=8cm]{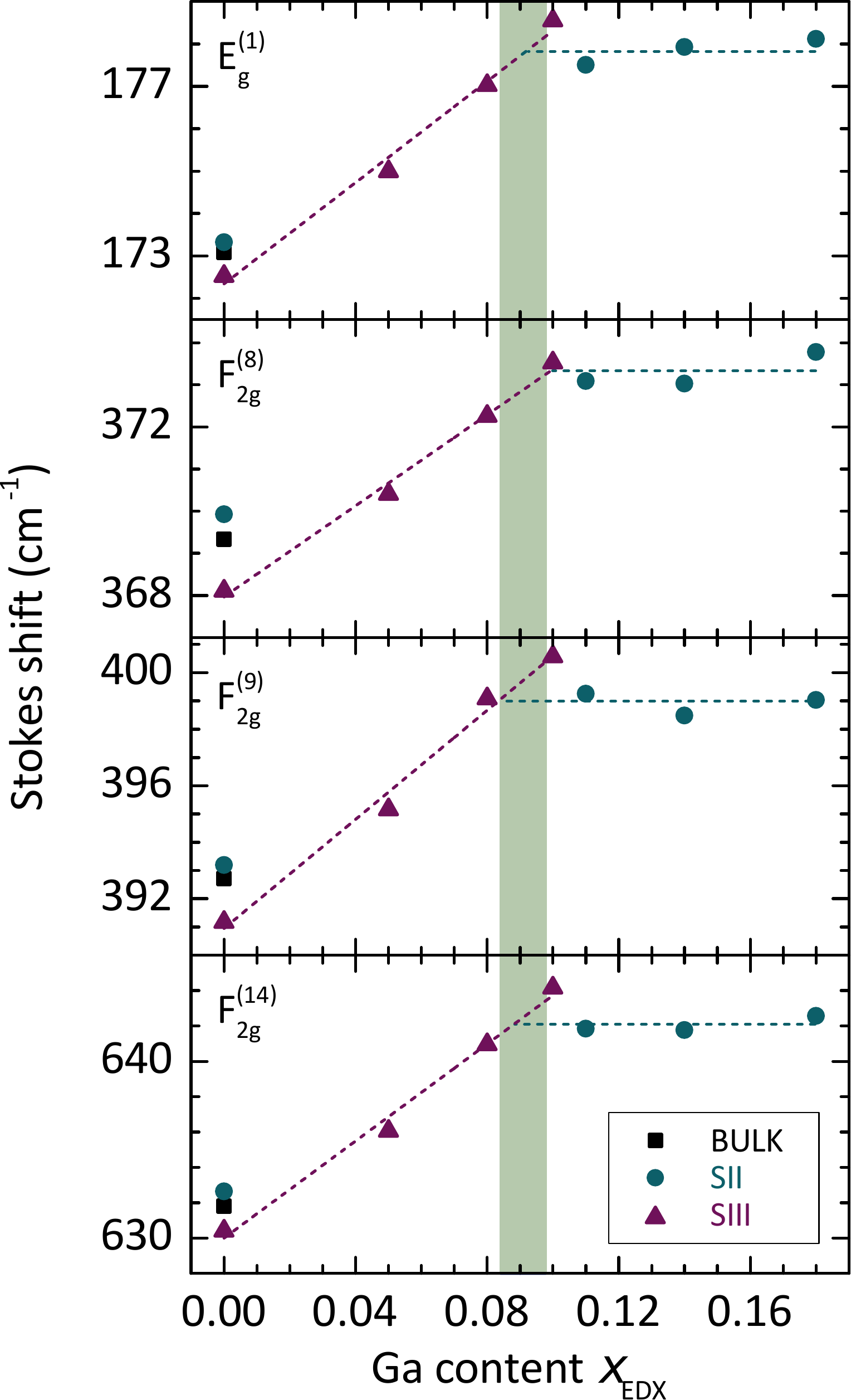}\centering}

\textcolor{black}{\caption{\textcolor{black}{Frequencies of several optical phonon modes for
the epitaxial films of sample series II as a function of the Ga content
determined by EDX along with the data for sample series III reported
in Ref.~\onlinecite{Feldl_2021}. The frequencies of a In}\protect\textsubscript{\textcolor{black}{2}}\textcolor{black}{O}\protect\textsubscript{\textcolor{black}{3}}\textcolor{black}{{}
bulk sample are shown for reference. Based on the observed linear
dependencies found for $x<0.10$, Raman phonon modes imply that even
though the samples of series II have nominally $x>0.10$ based on
EDX measurements, the effective Ga incorporation is approximately
$x=0.08-0.09$ on average, as indicated by the shaded area.\label{fig:Phonons} }}
}
\end{figure}

In \textcolor{black}{Ref.~\onlinecite{Feldl_2021}} a monotonous
blueshift with increasing Ga contents up to $x=0.10$ has been observed
for several optical phonon modes in MBE grown (In\textsubscript{1-\textit{x}}Ga\textsubscript{\textit{x}})\textsubscript{2}O\textsubscript{3}
films (sample series III), demonstrating that Raman spectroscopy can
be a suitable tool to determine the Ga incorporation in this alloy,
in the case of homogeneous Ga distribution. Figure~\ref{fig:Phonons}
displays the frequency of such phonon modes for the samples of series
II, which includes films with nominally $x>0.10$, along with the
data reported in \textcolor{black}{Ref.~\onlinecite{Feldl_2021}}.
The Raman spectra of all alloy films exhibit a typical fingerprint
of In\textsubscript{2}O\textsubscript{3} phonon modes with no indication
of additional phases (as in the high quality bulk crystal in Ref.~\textcolor{black}{\onlinecite{Galazka_2014}})
with no indication of additional phases.\textcolor{black}{{} The minor
discrepancies among the phonon frequencies of the }binary In\textsubscript{2}O\textsubscript{3}\textcolor{black}{{}
films of different sample series can be potentially attributed to
different degrees of lattice strain.} Strikingly, the frequencies
of all phonon modes remain constant for nominal Ga contents above
\textcolor{black}{$x>0.10$, with the }saturation values of the individual
phonon frequencies agreeing reasonably well with those obtained for
$x=0.10$.\citep{Feldl_2021} \textcolor{black}{The absence of a further
blueshift with increasing nominal Ga content is most likely related
to} the findings of the TEM investigations reporting a Ga incorporation
of approximately $x=0.08$ in the main ``matrix'' of the film and
inclusions with significantly higher Ga content for those films. Apparently,
the Raman spectra are dominated by scattering in the ``matrix''
of the films and not by the total Ga content of the films. The total
volume of the regions with extraordinary large Ga contents observed
by TEM is most likely too small to be detected in the Raman spectra.

\textcolor{black}{}
\begin{figure}[h]
\textcolor{black}{\includegraphics[width=8cm]{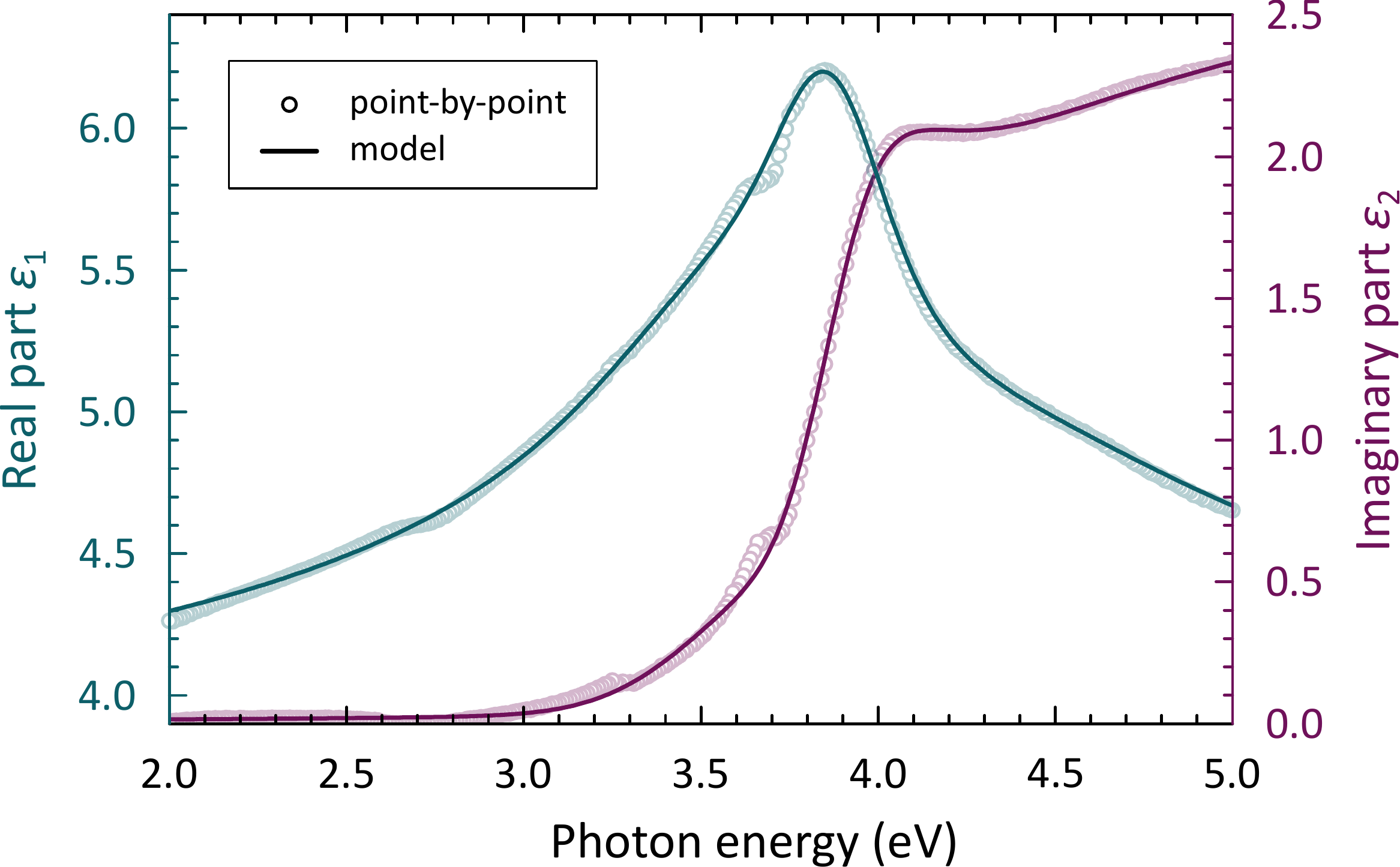}\centering}

\textcolor{black}{\caption{\textcolor{black}{Real, $\varepsilon_{1}$, and imaginary, $\varepsilon_{2}$,
parts of the dielectric functions for the }(In\protect\textsubscript{0.89}Ga\protect\textsubscript{0.11})\protect\textsubscript{2}O\protect\textsubscript{3}\textcolor{black}{{}
film (II.b) determined by spectroscopic ellipsometry. Both, the point-by-point
fitted result (open circles) and the dielectric function as described
by a model function (lines) are shown.\label{fig:DF}}}
}
\end{figure}

\textcolor{black}{}
\begin{figure}[h]
\textcolor{black}{\includegraphics[width=8cm]{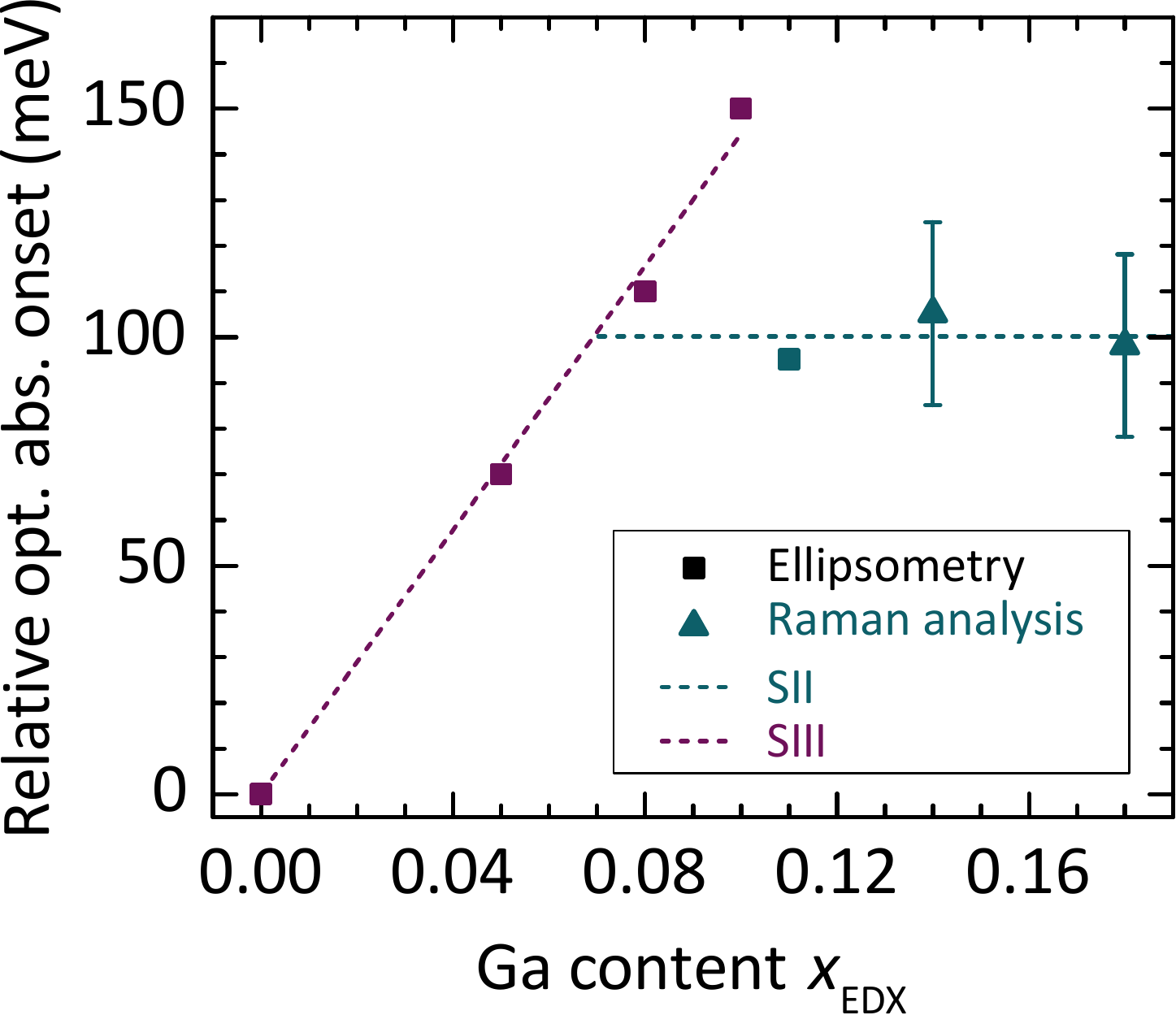}\centering}

\textcolor{black}{\caption{\textcolor{black}{Shift of optical absorption onset of alloy films
with $x=0.14$ (II.c) and $x=0.18$ (II.d) with respect to that of
In}\protect\textsubscript{\textcolor{black}{2}}\textcolor{black}{O}\protect\textsubscript{\textcolor{black}{3}}\textcolor{black}{{}
determined by the analysis of the YSZ substrate signal in Raman spectra
(triangles). The corresponding blueshift of the alloy film with $x=0.11$
and for sample series III (reported in Ref.~\onlinecite{Feldl_2021})
are based on the analysis of the dielectric functions, such as this
shown in Fig.~\ref{fig:DF} for II.b. The dashed lines are linear
fits of the data of sample series II and III (teal and purple, respectively)
and intended as guides to the eye.\label{fig:Raman_3}}}
}
\end{figure}

Spectroscopic ellipsometry is demonstrated in \textcolor{black}{Ref.~\onlinecite{Feldl_2021}}
to be a further spectroscopic tool to study the incorporation of Ga
in (In\textsubscript{1-\textit{x}}Ga\textsubscript{\textit{x}})\textsubscript{2}O\textsubscript{3}
alloy films. The onset of optical absorption has been shown to exhibit
a pronounced blueshift with increasing Ga content for the samples
of series III. Regarding sample series II, the dielectric function
of the sample with $x=0.11$ is shown in Fig.~\ref{fig:DF}. A comparison
between the dielectric functions of the samples of series II and pure
In\textsubscript{2}O\textsubscript{3} can be found in the recent
work of \citet{Feldl_2021}. The observed blueshift of the absorption
onset by 95~meV induced by al\textcolor{black}{loying is somewhat
lower than expected based on the results reported in Ref.~\onlinecite{Feldl_2021},
presumably due to the different growth conditions.} Note that many-body
corrections such as the Burstein-Moss effect and band gap renormalization
can be neglected for the investigated sample,\citep{feneberg_2016}
which has been annealed in oxygen (rapid thermal annealing at a final
temperature of $800\,\mathrm{{^\circ}C}$ at atmospheric pressure
for 60~s).\textcolor{black}{{} Due to rough surface morphologies (see
Fig.~\ref{fig:AFM_all})}, samples with $x>0.11$ could not be analyzed
by spectroscopic ellipsometry. Instead, we developed a method to determine
the blueshift of the absorption onset in such alloy films by Raman
spectroscopy. Our approach is based on the fact that the relative
contribution of Raman scattering in the YSZ substrate is directly
correlated with the optical absorption in the (In\textsubscript{1-\textit{x}}Ga\textsubscript{\textit{x}})\textsubscript{2}O\textsubscript{3}
alloy films. Utilizing optical excitation at 3.81~eV (close to the
onset of absorption), the intensity of the YSZ substrate signal in
Raman spectra can be used to determine the alloying-induced blueshift
of the optical absorption onset in those rough samples. A detailed
description of our approach and a verification of its validity are
presented in the Supplemental Material. The obtained relative optical
absorption onsets for $x=0.14$ and $0.18$ are shown in Fig.~\ref{fig:Raman_3}
along with the value determined by spectroscopic ellipsometry for
$x=0.11$ (see Fig.~\ref{fig:DF}). In accordance with the result
obtained for the frequencies of optical phonons, no clear dependence
of the absorption onset on the nominal Ga content is found for $x>0.11$
within the limits of accuracy (uncertainty of about 20~meV). Consequently,
the saturation of the blueshifts found for both the phonon frequencies
and the optical absorption onset confirm the phase separation observed
by TEM for alloy films with nominal Ga contents above about 10\,\%.

\textcolor{black}{}
\begin{figure}[h]
\textcolor{black}{\includegraphics[width=7cm]{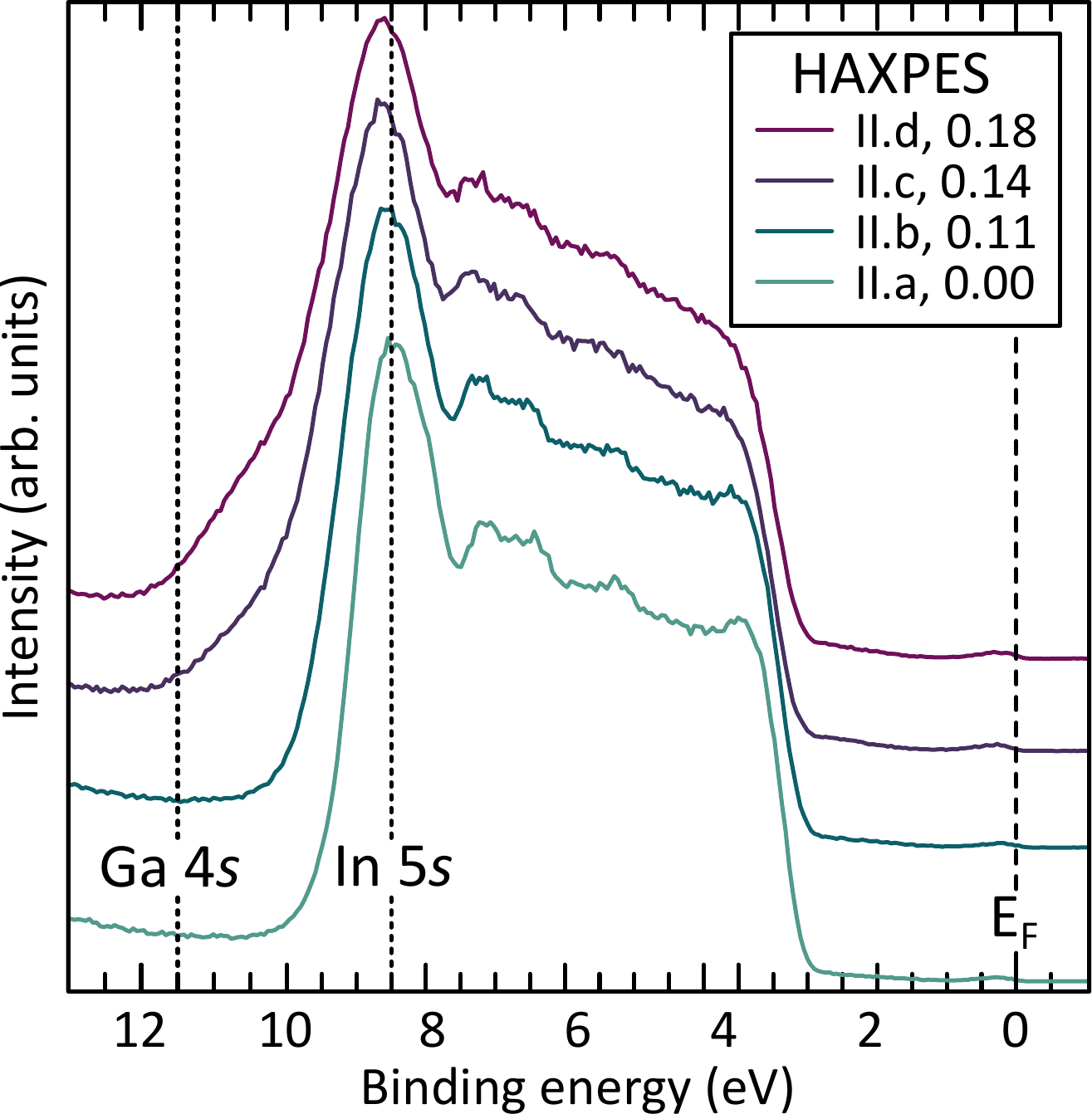}\centering}

\textcolor{black}{\caption{\textcolor{black}{HAXPES valence band spectra of }(In\protect\textsubscript{1-\textit{x}}Ga\protect\textsubscript{\textit{x}})\protect\textsubscript{2}O\protect\textsubscript{3}\textcolor{black}{{}
with nominal Ga contents of $x=0.00$, 0.11, 0.14, and 0.18 (sample
series II).\label{fig:HAXPES_VB}}}
}
\end{figure}

\textcolor{black}{}
\begin{figure}[h]
\textcolor{black}{\includegraphics[width=6cm]{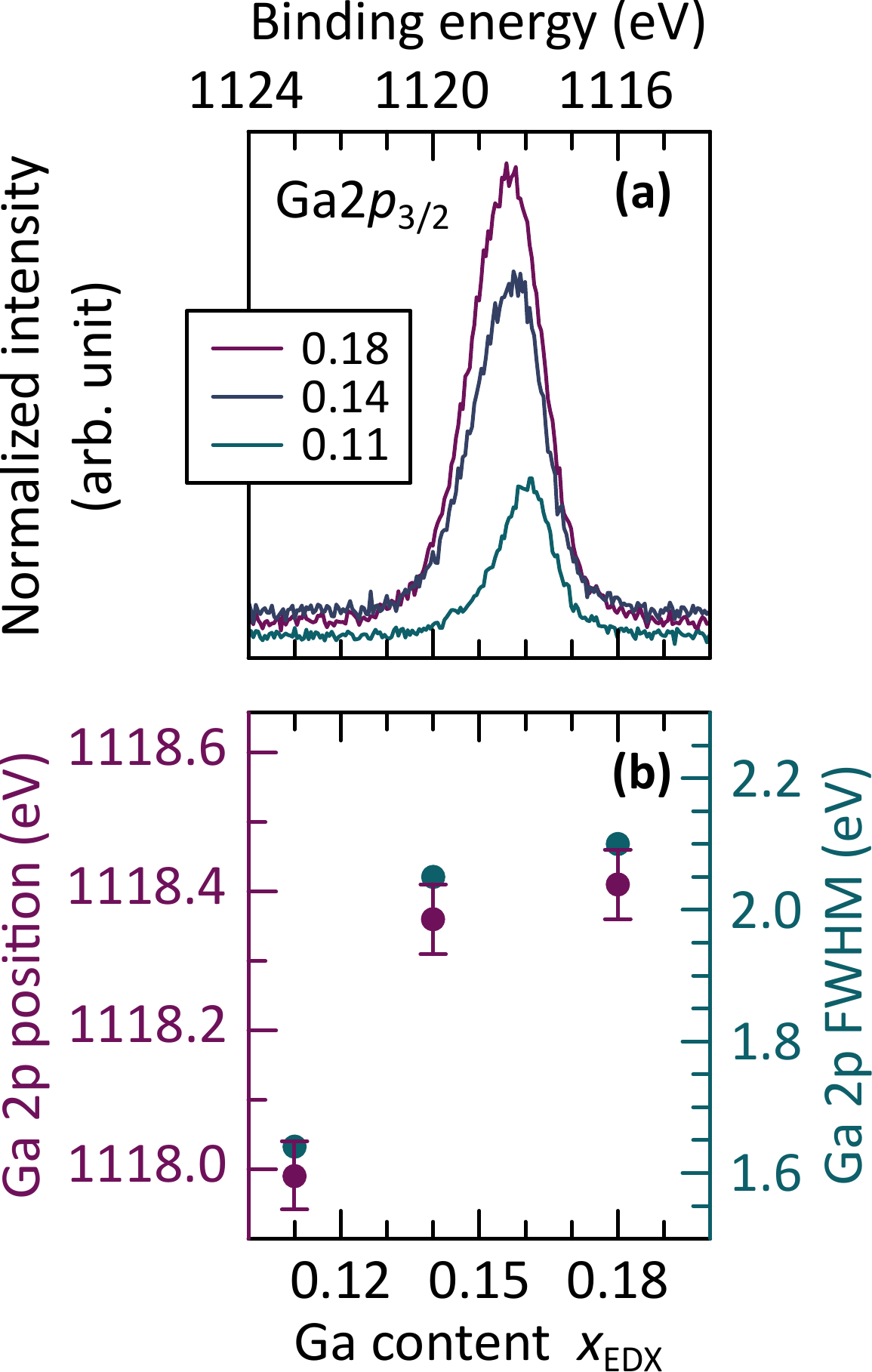}\centering}

\textcolor{black}{\caption{\textcolor{black}{(a) HAXPES Ga 2}\textit{\textcolor{black}{p}}\textcolor{black}{{}
core level binding energies and peak shapes and (b) comparison of
peak position of all alloy layers of series II and corresponding FWHM
as a function of Ga content.\label{fig:HAXPES_Ga2p}}}
}
\end{figure}

\textcolor{black}{Finally, in order to investigate chemical bonding
and electronic states of the alloy, the films of series II have been
investigated by HAXPES. The valence band structure of three of those
(II.a, b, and d) can be seen in Fig.~\ref{fig:HAXPES_VB}. }For In\textsubscript{2}O\textsubscript{3}
and Ga\textsubscript{2}O\textsubscript{3}, features around 8.5 and
11.5~eV are expected, corresponding to the In 5\textit{s} and Ga
4\textit{s} orbitals, respectively.\textcolor{black}{{} As seen in Fig.~\ref{fig:HAXPES_VB},
the addition of Ga shifts the spectrum to higher binding energies.
For the II.d. film with $x=0.18$, in particular, an additional, distinct
feature can be observed at the high-binding-energy end. This could
be attributed to the inclusions with very high Ga contents having
distinctly different valence band structure.} Charge distributions
of the Ga bonding state were confirmed in the Ga 2\textit{p} spectra
for all Ga-containing samples of series II, as shown in Fig.~\textcolor{black}{\ref{fig:HAXPES_Ga2p}}.
Both the peak position and FWHM values of Ga 2\textit{p} core level
change significantly for $x>0.11$.\textcolor{red}{{} }\textcolor{black}{The
saturation of core level position likely reflects the ``matrix''
of the film, similar to the saturation observed in the Raman line
frequencies shift and optical band gap width.}

\textcolor{black}{Hence, the investigations and analysis of the} (In\textsubscript{1-\textit{x}}Ga\textsubscript{\textit{x}})\textsubscript{2}O\textsubscript{3}\textcolor{black}{{}
alloy films using spectroscopic techniques, along with the combination
of TEM and EDX data, revealed that XRD results inferring a homogeneous
incorporation of the Ga cations up to $x=0.18$ can be misleading.}

\section{Conclusion}

The heteroepitaxial growth of (In\textsubscript{1-\textit{x}}Ga\textsubscript{\textit{x}})\textsubscript{2}O\textsubscript{3}
on YSZ is enhanced by the employment of a thin pure \textcolor{black}{In}\textsubscript{\textcolor{black}{2}}\textcolor{black}{O}\textsubscript{\textcolor{black}{3}}\textcolor{black}{{}
buffer layer between the alloy film and the substrate, which provides
a better-matched substrate in terms of wetting, chemistry, and lattice
parameter. This has been demonstrated in terms of small full widths
at half maximum of x-ray diffraction} rocking curves, which indicate
higher film crystallinity, as well as smoother film surfaces. Both
x-ray diffraction and transmission electron microscopy investigations
confirm the cubic phase purity and single-crystallinity of the films
up to nominal Ga contents of $x=0.18$. Moreover, the lattice parameters
of the films measured by x-ray diffraction reciprocal space mapping
obey Vegard's law, assuming the Ga contents measured by energy dispersive
x-ray spectroscopy. These findings would imply a homogeneous incorporation
of Ga in the alloy films.

However, closer inspection of dark features in transmission electron
micrographs of the samples with the highest Ga compositions and local
energy dispersive x-ray spectroscopy measurements reveal Ga containing
inclusions in the films with Ga contents as high as $x=0.50$, while
the remaining film, the \textcolor{black}{``matrix''}, exhibits
an average Ga incorporation of approximately $x=0.08$. Nevertheless,
both of these regions preserve the cubic bixbyite structure. \textcolor{black}{The
analysis of the} (In\textsubscript{1-\textit{x}}Ga\textsubscript{\textit{x}})\textsubscript{2}O\textsubscript{3}\textcolor{black}{{}
alloy films by Raman spectroscopy, spectroscopic ellipsometry, and
hard x-ray photoelectron spectroscopy corroborates this finding. They
show systematic shifts up to an average composition of approximately}
$x=0.10$\textcolor{black}{{} and saturation for further increasing
$x$, in which case their results largely reflect the properties of
the ``matrix'' rather than that of the average material, i.e., both
the ``matrix'' and inclusions with high $x$.}

\textcolor{black}{Thus, the observations of x-ray diffraction and
reciprocal space maps can be misleading. Since the changes in lattice
parameter\textemdash larger for In}\textsubscript{\textcolor{black}{2}}\textcolor{black}{O}\textsubscript{\textcolor{black}{3}}\textcolor{black}{{}
and smaller with increasing $x$ for (In}\textsubscript{\textcolor{black}{1-}\textit{\textcolor{black}{x}}}\textcolor{black}{Ga}\textsubscript{\textit{\textcolor{black}{x}}}\textcolor{black}{)}\textsubscript{\textcolor{black}{2}}\textcolor{black}{O}\textsubscript{\textcolor{black}{3}}\textcolor{black}{\textemdash happen
within an elastic medium, relaxation between areas with different
Ga contents takes place. An effect of ``averaging'' can be observed
in this respect. On the contrary, this averaging effect is not visible
in local microscopy investigations and spectroscopic methods that
are not largely affected by strain and, thus, the larger volume of
the film ``matrix'' is reflected within these, rather than the inclusions
with high $x$.}

\section*{Acknowledgment}

We would like to\textcolor{black}{{} thank Duc Van Dinh and Thomas Teubner
for} critically reading this manuscript, as well as Uwe Jahn for help
with the EDX investigations of the Ga content of the films, Anne-Kathrin
Bluhm for the cross-sectional SEM images for film thickness characterization,
and Hans-Peter Schönherr and Carsten Stemmler for technical assistance
with the MBE system. This study was performed in the framework of
GraFOx, a Leibniz-ScienceCampus partially funded by the Leibniz Association.
We are also grateful to HiSOR, Hiroshima University, and JAEA/SPring-8
for the development of HAXPES at BL15XU of SPring-8. The HAXPES measurements
were performed under the approval of the NIMS Synchrotron X-ray Station
(Proposal No. 2019B4602).

\bibliographystyle{apsrev4-1}
\bibliography{IGO_literature}

\end{document}


\title{Supplemental material: Molecular beam epitaxy of single-crystalline
bixbyite (In\textsubscript{1-\textit{x}}Ga\textsubscript{\textit{x}})\textsubscript{2}O\textsubscript{3}
films (\textit{x}\ensuremath{\le}0.18): Structural properties and
consequences of compositional inhomogeneity }
\author{Alexandra Papadogianni}
\affiliation{Paul-Drude-Institut für Festkörperelektronik, Leibniz-Institut im
Forschungsverbund Berlin e.V., Hausvogteiplatz 5\textendash 7, 10117
Berlin, Germany}
\author{Charlotte Wouters}
\affiliation{Leibniz-Institut für Kristallzüchtung, Max-Born-Str. 2, 12489 Berlin,
Germany}
\author{Robert Schewski}
\affiliation{Leibniz-Institut für Kristallzüchtung, Max-Born-Str. 2, 12489 Berlin,
Germany}
\author{Johannes Feldl}
\affiliation{Paul-Drude-Institut für Festkörperelektronik, Leibniz-Institut im
Forschungsverbund Berlin e.V., Hausvogteiplatz 5\textendash 7, 10117
Berlin, Germany}
\author{Jonas Lähnemann}
\affiliation{Paul-Drude-Institut für Festkörperelektronik, Leibniz-Institut im
Forschungsverbund Berlin e.V., Hausvogteiplatz 5\textendash 7, 10117
Berlin, Germany}
\author{\textcolor{black}{Takahiro Nagata}}
\affiliation{\textcolor{black}{National Institute for Materials Science, 1-1 Namiki
Tsukuba, 305-0044 Ibaraki, Japan}}
\author{Elias Kluth}
\affiliation{Institut für Experimentelle Physik, Otto-von-Guericke-Universität
Magdeburg, Universitätsplatz 2, 39106 Magdeburg, Germany}
\author{Martin Feneberg}
\affiliation{Institut für Experimentelle Physik, Otto-von-Guericke-Universität
Magdeburg, Universitätsplatz 2, 39106 Magdeburg, Germany}
\author{Rüdiger Goldhahn}
\affiliation{Institut für Experimentelle Physik, Otto-von-Guericke-Universität
Magdeburg, Universitätsplatz 2, 39106 Magdeburg, Germany}
\author{Manfred Ramsteiner}
\affiliation{Paul-Drude-Institut für Festkörperelektronik, Leibniz-Institut im
Forschungsverbund Berlin e.V., Hausvogteiplatz 5\textendash 7, 10117
Berlin, Germany}
\author{Martin Albrecht}
\affiliation{Leibniz-Institut für Kristallzüchtung, Max-Born-Str. 2, 12489 Berlin,
Germany}
\author{Oliver Bierwagen}
\affiliation{Paul-Drude-Institut für Festkörperelektronik, Leibniz-Institut im
Forschungsverbund Berlin e.V., Hausvogteiplatz 5\textendash 7, 10117
Berlin, Germany}
\maketitle

\subsection{Transmission electron microscopy}

\begin{figure}[H]
\textcolor{black}{\includegraphics[width=8cm]{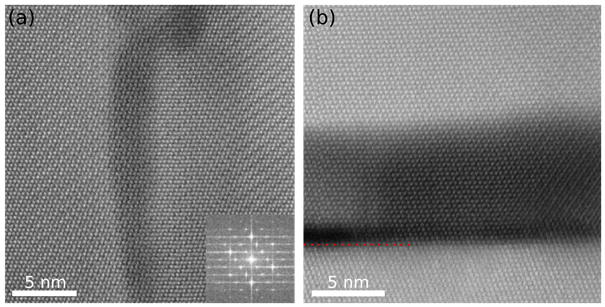}\centering}

\textcolor{black}{\caption{HAADF-STEM images of (a) a vertically travelling void in the pure
In\protect\textsubscript{2}O\protect\textsubscript{3} film and (b)
an interface void in the film with $x=0.11$.\label{fig:SuppTEM_Fig1}}
}
\end{figure}

\textcolor{black}{Additional }HAADF-STEM \textcolor{black}{images
showing voids in the bulk of the pure }In\textsubscript{2}O\textsubscript{3}\textcolor{black}{film
(II.a) and at the interface between the }(In\textsubscript{0.89}Ga\textsubscript{0.11})\textsubscript{2}O\textsubscript{3}\textcolor{black}{{}
film (II.b) and the YSZ substrate are shown in Figure~\ref{fig:SuppTEM_Fig1}.
Both the UID and $x=0.11$ films show dark features close to the substrate,
which are caused by voids, i.e., regions of empty space in the material.
This void formation can be explained in terms of weak bonding between
the In}\textsubscript{\textcolor{black}{2}}\textcolor{black}{O}\textsubscript{\textcolor{black}{3}}\textcolor{black}{{}
buffer layer and the YSZ substrate and 3D island formation due to
faster growth than on the substrate\citep{Bierwagen_faceting_2016,Bierwagen_JAP_2010}
that coalesce and leave voids behind. }The cubic bixbyite structure
is not interrupted.

\begin{figure}[H]
\includegraphics[width=8cm]{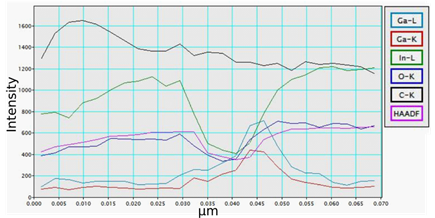}\centering

\caption{X-ray peak intensities of the different elements for a EDX line scan
across one of the dark intensity features in the film with $x=0.14$.\label{fig:SuppTEM_Fig2}}
\end{figure}

\textcolor{black}{Figure~\ref{fig:SuppTEM_Fig2} depicts EDX line
scans showing a dip in O content coinciding with the Ga increase,
indicating further void formation.}

\textcolor{black}{}

\subsection{Determination of optical absorption onset by Raman spectroscopy}

\begin{figure}[H]
\includegraphics[width=8cm]{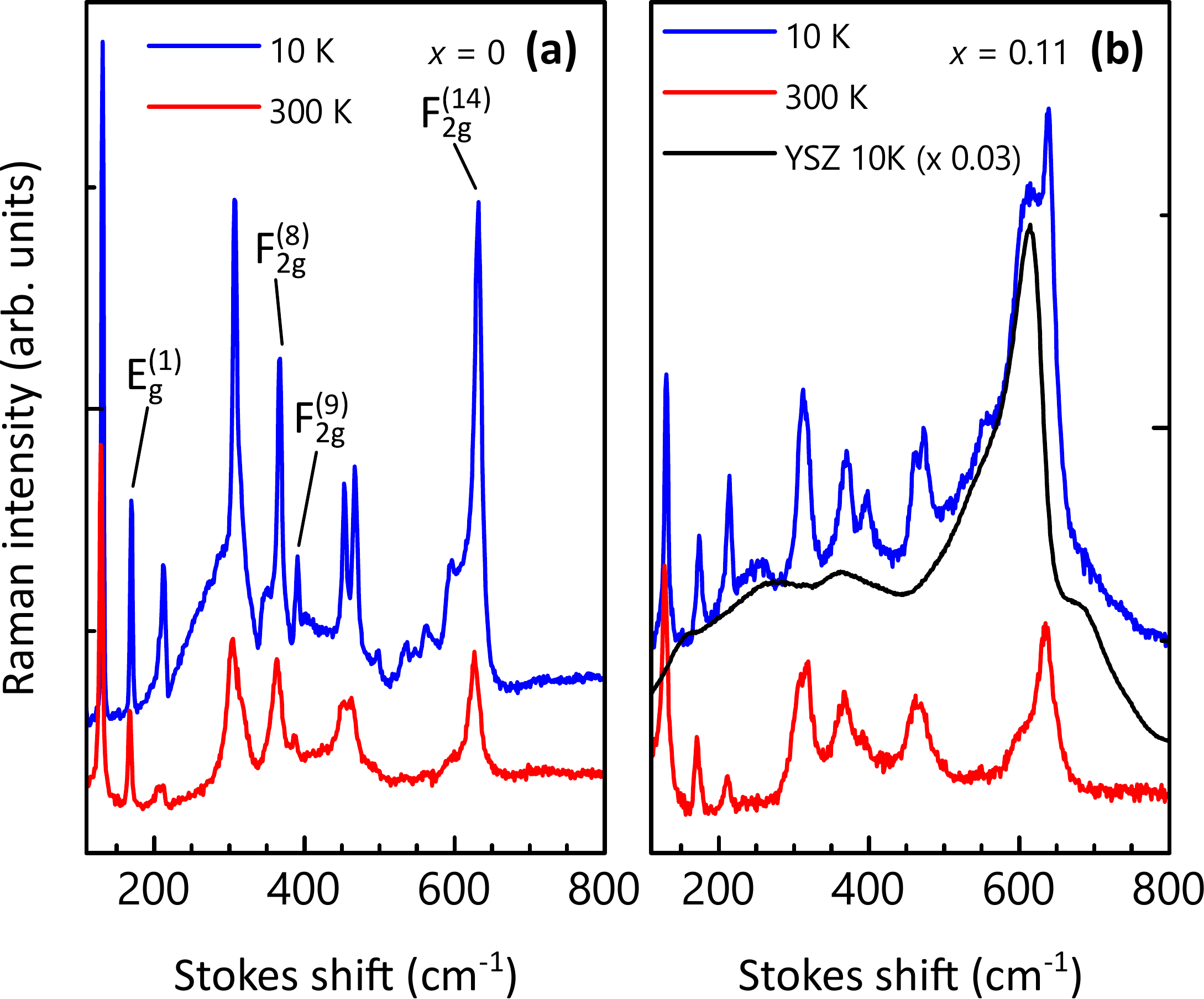}\centering

\caption{Raman spectra of \textcolor{black}{In}\protect\textsubscript{\textcolor{black}{2}}\textcolor{black}{O}\protect\textsubscript{\textcolor{black}{3}}
(a) and (In\protect\textsubscript{0.89}Ga\protect\textsubscript{0.11})\protect\textsubscript{2}O\protect\textsubscript{3}
(b) films measured at room temperature and at 10~K. In (b) the Raman
spectrum of bulk YSZ (111) is shown for comparison.\label{fig:SuppRaman_Fig1}}
\end{figure}

\begin{figure}
\includegraphics[width=8cm]{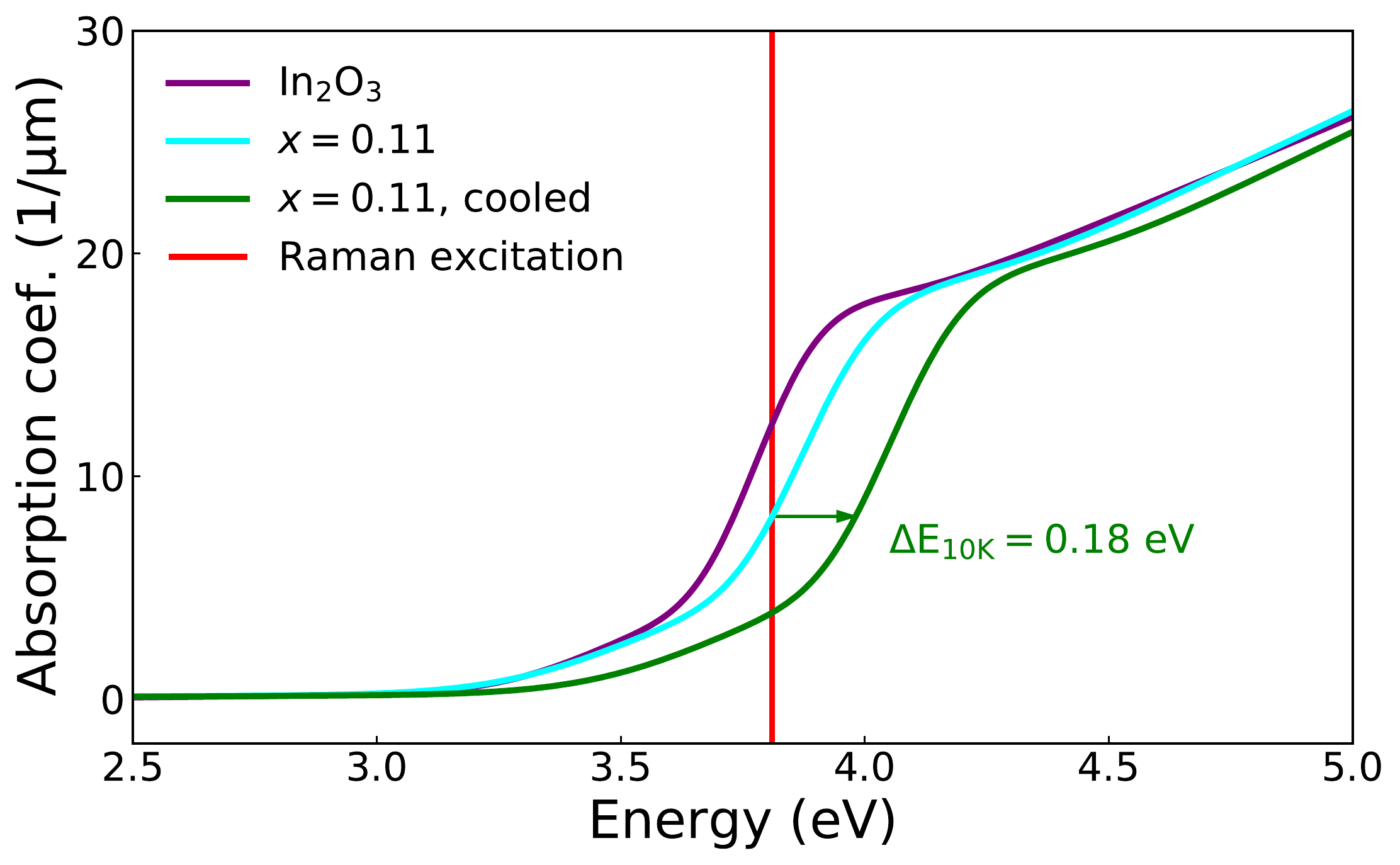}\centering

\caption{Absorption spectrum of \textcolor{black}{In}\protect\textsubscript{\textcolor{black}{2}}\textcolor{black}{O}\protect\textsubscript{\textcolor{black}{3}}
and (In\protect\textsubscript{0.89}Ga\protect\textsubscript{0.11})\protect\textsubscript{2}O\protect\textsubscript{3}
films derived from the model of their dielectric functions measured
by spectroscopic ellipsometry. The spectrum of the (In\protect\textsubscript{0.89}Ga\protect\textsubscript{0.11})\protect\textsubscript{2}O\protect\textsubscript{3}
film is shown in addition with a rigid energy shift $\Delta E$ according
to the bandgap shift between room temperature and 10~K. The vertical
line is drawn at the photon energy used for the excitation of our
Raman experiments.\label{fig:SuppRaman_Fig2}}
\end{figure}

Our approach for the determination of the optical absorption onset
is based on the fact that the intensity of a spectral feature from
the YSZ substrate in the Raman spectrum of an alloy sample depends
on the optical absorption in the epitaxial film. Figure~\ref{fig:SuppRaman_Fig1}
displays Raman spectra of \textcolor{black}{In}\textsubscript{\textcolor{black}{2}}\textcolor{black}{O}\textsubscript{\textcolor{black}{3}}
and (In\textsubscript{0.89}Ga\textsubscript{0.11})\textsubscript{2}O\textsubscript{3}
on YSZ substrates for excitation close to the onset of optical absorption
at 3.81~eV. All spectra exhibit the characteristic phonon lines of
epitaxial bixbyite films.\textcolor{red}{\citep{kranert_2014a}} However,
as a consequence of the band gap widening, the spectra of the alloy
film contain an additional spectral feature around $620\,\mathrm{cm^{-1}}$
originating from Raman scattering at the YSZ substrate. This contribution
becomes even more pronounced at low temperatures. The dependence of
the YSZ Raman signal intensity on the optical absorption onset is
illustrated in Fig.~\ref{fig:SuppRaman_Fig2} which displays the
absorption coefficients of \textcolor{black}{In}\textsubscript{\textcolor{black}{2}}\textcolor{black}{O}\textsubscript{\textcolor{black}{3}}
and (In\textsubscript{0.89}Ga\textsubscript{0.11})\textsubscript{2}O\textsubscript{3}
films in the vicinity of the absorption onset as derived from the
dielectric functions. At the photon energy used for optical excitation
of the Raman spectra (3.81~eV), the absorption in the \textcolor{black}{In}\textsubscript{\textcolor{black}{2}}\textcolor{black}{O}\textsubscript{\textcolor{black}{3}}
fil\textcolor{black}{m is sufficiently strong to suppress significant
Raman scattering in the substrate already for film thicknesses of
about 300~nm (see Table~1 o}f main article). The absorption in the
alloy film, on the other hand, is strongly reduced leading to a considerable
contribution of Raman scattering originating from the substrate. At
low temperatures, the (In\textsubscript{0.89}Ga\textsubscript{0.11})\textsubscript{2}O\textsubscript{3}
film becomes even more transparent due to the further band gap widening
and the YSZ Raman signal even more pronounced (see Fig.~\ref{fig:SuppRaman_Fig1}).

The intensity of the substrate Raman signal is given by the exponential
decay of both the incoming ($I_{\mathrm{0}}$) and outgoing ($I_{\mathrm{S}}$)
light:\citealp{richter_1976}

\begin{equation}
I_{\mathrm{S}}=I_{\mathrm{0}}\exp[-2a_{\mathrm{R}}d]\label{eq:EQScatterInt}
\end{equation}
 with

\begin{equation}
\alpha_{\mathrm{R}}=\alpha(h\nu_{\mathrm{R}}-\Delta E)\label{eq:EQEffAbsCoeff}
\end{equation}

\begin{equation}
\alpha=\frac{4\pi k}{\lambda}\label{eq:EQAbsCoeff}
\end{equation}

\begin{equation}
k=\sqrt{\frac{\left(\epsilon_{1}^{2}+\epsilon_{2}^{2}\right)^{\frac{1}{2}}-\epsilon_{1}}{2}}\label{eq:EQk}
\end{equation}
.

Here $d$ is the thickness of the (In\textsubscript{1-\textit{x}}Ga\textsubscript{\textit{x}})\textsubscript{2}O\textsubscript{3}
film, $k$ the complex part of the refractive index, $\lambda$ the
photon wavelength and $\epsilon_{1}$ and $\epsilon_{2}$ the real
and imaginary parts of the dielectric function, respectively. The
slightly different photon energies of the incoming ($h\nu_{\mathrm{L}}$)
and outgoing ($h\nu_{\mathrm{S}}$) light are taken into account by
considering the absorption $\alpha_{\mathrm{R}}$ at the average photon
energy $h\nu_{\mathrm{R}}=(h\nu_{\mathrm{L}}+h\nu_{\mathrm{S}})=3.77\,\mathrm{eV}$.
For our model, we use the absorption spectrum $\alpha(h\nu)$ of the
(In\textsubscript{0.89}Ga\textsubscript{0.11})\textsubscript{2}O\textsubscript{3}
film shown in Fig.~\ref{fig:SuppRaman_Fig1}. The absorption spectra
of other alloy films with larger Ga contents are approximated by a
rigid energy shift $\Delta E$ in order to simulate the blueshift
of the optical absorption onset.\textcolor{teal}{{} }\textcolor{black}{For
the analysis of alloy samples with $x>0.11$, the intensity of the
substrate signal $I_{\mathrm{S}}$ is normalized to that of the (In}\textsubscript{\textcolor{black}{0.89}}\textcolor{black}{Ga}\textsubscript{\textcolor{black}{0.11}}\textcolor{black}{)}\textsubscript{\textcolor{black}{2}}\textcolor{black}{O}\textsubscript{\textcolor{black}{3}}\textcolor{black}{{}
reference sample ($I_{\mathrm{S,0}}$) with $\Delta E=0$. With the
approximation of a fixed prefactor $I_{0}$, the obtained intensity
ratio is given by}

\textcolor{black}{
\begin{equation}
\frac{I_{\mathrm{S}}}{I_{\mathrm{S,0}}}=\frac{\exp\left[-2\alpha(h\nu_{\mathrm{R}}\text{\textendash}\Delta E)d\right]}{\exp\left[-2\alpha(h\nu_{\mathrm{R}})d_{0}\right]}\label{eq:intensity_ratio}
\end{equation}
}

\textcolor{black}{This equation can be fulfilled only by a specific
shift $\Delta E$ in the optical absorption onset for the film with
larger Ga content. Consequently, the measured values of $\frac{I_{\mathrm{S}}}{I_{\mathrm{S,0}}}$
are used to determine $\Delta E$, the only unknown parameter in Eq.~\ref{eq:intensity_ratio}.
For the normalization by $I_{\mathrm{S,0}}$, the Raman spectrum of
the (In}\textsubscript{\textcolor{black}{0.89}}\textcolor{black}{Ga}\textsubscript{\textcolor{black}{0.11}}\textcolor{black}{)}\textsubscript{\textcolor{black}{2}}\textcolor{black}{O}\textsubscript{\textcolor{black}{3}}\textcolor{black}{{}
reference sample has been measured under exactly the same conditions
along with each measurement on an alloy sample with larger Ga content.}

\begin{figure}
\includegraphics[width=8cm]{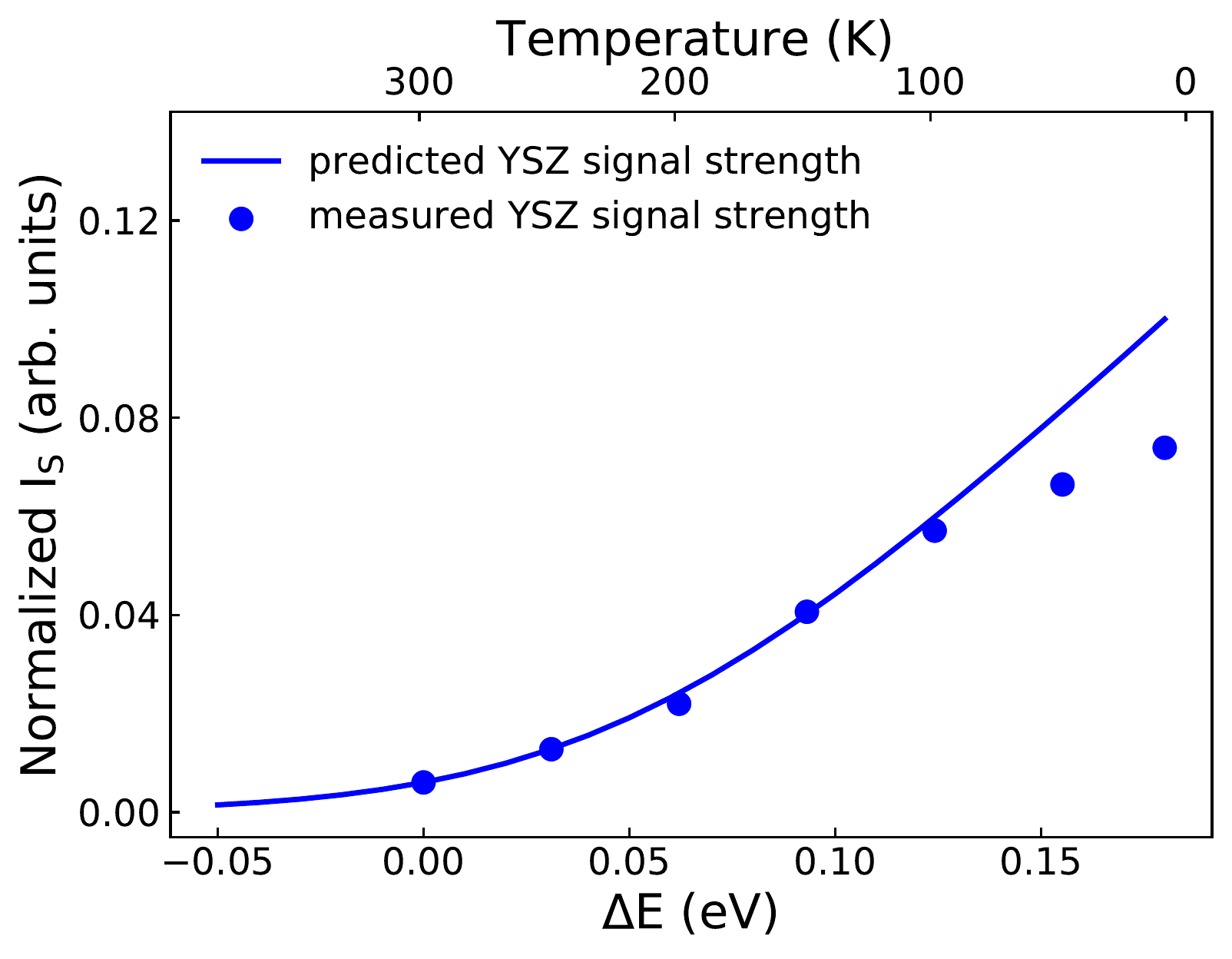}\centering

\caption{Measured and calculated intensity of the Raman signal from the YSZ
substrate for a (In\protect\textsubscript{0.89}Ga\protect\textsubscript{0.11})\protect\textsubscript{2}O\protect\textsubscript{3}
sample as a function of the temperature dependent energy shift $\Delta E(T)$.\label{fig:SuppRaman_Fig3}}
\end{figure}

\textcolor{black}{In order to verify the validity of our approach,
we performed Raman measurements for which the optical absortption
onset of the (In}\textsubscript{\textcolor{black}{0.89}}\textcolor{black}{Ga}\textsubscript{\textcolor{black}{0.11}}\textcolor{black}{)}\textsubscript{\textcolor{black}{2}}\textcolor{black}{O}\textsubscript{\textcolor{black}{3}}\textcolor{black}{{}
film, and thus $\Delta E$, was tuned by varying the sample temperature
($T$). Thereby, $\Delta E$(T) was assumed to be identical to the
temperature dependent shift of the fundamental band gap in In}\textsubscript{\textcolor{black}{2}}\textcolor{black}{O}\textsubscript{\textcolor{black}{3}}\textcolor{black}{.\citep{Irmscher_10.1002/pssa.201330184}
For normalization purposes, the Raman signal of a YSZ (111) bulk sample
has been measured under exactly the same experimental conditions.
Our simulation according to Eq.~\ref{eq:EQScatterInt} is shown in
Fig.~\ref{fig:SuppRaman_Fig3} together with the measured substrate
intensity as a function of the temperature dependent shift $\Delta E$.
The good agreement between simulation and experiment demonstrates
that our approach can be used to determine the shift of the optical
absorption onset $\Delta E(x)$ at room temperature for alloy films
with $x>0.11$.}

Our approach is limited to film thicknesses which are sufficiently
transparent at $h\nu=3.8\,\mathrm{eV}$ to detect a significant Raman
signal from the YSZ substrate. For our model the spectral shape of
the absorption onset has to remain essentially unaltered for different
alloy compositions. Furthermore, optical interference effects due
to multiple reflections inside the alloy films have to be negligible.

\date{\today}

\bibliographystyle{apsrev4-1}
\bibliography{IGO_literature}